\documentclass[pra,twocolumn,superscriptaddress,amsmath,amssymb,aps,floatfix]{revtex4-2}
\usepackage{placeins}
\usepackage{graphicx}
\usepackage{dcolumn}
\usepackage{bm}
\usepackage{amsmath,amssymb}
\usepackage{graphics}
\usepackage{amsfonts}
\usepackage{epsfig}
\usepackage{float}

\usepackage{adjustbox}
\usepackage{rotating}
\usepackage{float}
\usepackage{dblfloatfix}
\usepackage{blindtext}
\usepackage{multirow}
\usepackage{graphicx}
\usepackage{dcolumn}
\usepackage{bbold}
\usepackage{bm}
\usepackage{amsmath}
\usepackage{import}
\usepackage{physics}
\usepackage{verbatim}
\usepackage{listings}
\usepackage{xcolor}

\usepackage{multirow}

\usepackage{caption}
\usepackage{subcaption}
\usepackage{algpseudocode}
\usepackage[colorlinks=true,allcolors=blue]{hyperref}
\usepackage{cleveref}
\begin{document}
\title{\textcolor{black}{Quantum cryptography compatible with noisy intermediate-scale quantum devices based on Parrondo dynamics in discrete-time quantum walks}}
\author{Aditi Rath}
\email{aditi.rath@niser.ac.in}
\author{Dinesh Kumar Panda}
\email{dineshkumar.quantum@gmail.com}
\author{Colin Benjamin}
\email{colin.nano@gmail.com}
\affiliation{School of Physical Sciences, National Institute of Science Education and Research, Bhubaneswar, Jatni 752050, India}
\affiliation{Homi Bhabha National Institute, Training School Complex, Anushaktinagar, Mumbai
400094, India}
\begin{abstract}
Compatibility with noisy intermediate-scale quantum (NISQ) devices is crucial for the realistic implementation of quantum cryptographic protocols. We investigate a cryptographic scheme based on discrete-time quantum walks (DTQWs) on cyclic graphs that exploits Parrondo dynamics, wherein periodic evolution emerges from a deterministic sequence of individually chaotic coin operators. We construct an explicit quantum circuit realization tailored to NISQ architectures and analyze its performance through numerical simulations in Qiskit under both ideal and noisy conditions. Protocol performance is quantified using probability distributions, Hellinger fidelity, and total variation distance. To assess security at the circuit level, we model intercept–resend and man-in-the-middle attacks and evaluate the resulting quantum bit error rate. In the absence of adversarial intervention, the protocol enables reliable message recovery, whereas eavesdropping induces characteristic disturbances that disrupt the periodic reconstruction mechanism. We further examine hardware feasibility on contemporary NISQ processors, specifically $ibm\_torino$, incorporating qubit connectivity and state-transfer constraints into the circuit design. Our analysis demonstrates that communication between spatially separated logical modules increases circuit depth via SWAP operations, leading to cumulative noise effects. By exploring hybrid state-transfer strategies, we show that qubit selection and connectivity play a decisive role in determining fidelity and overall protocol performance, highlighting hardware-dependent trade-offs in NISQ implementations.
\end{abstract}
\maketitle
\newpage
\twocolumngrid
\section{Introduction} \label{S1}
The BB84 quantum key distribution (QKD) protocol~\cite{bb84} marked the beginning of quantum cryptography research, which was greatly bolstered by the E91 entanglement-based protocol~\cite{Ekert}. Together, these pioneering schemes established that the fundamental principles of quantum mechanics, such as the impossibility of perfect state cloning~\cite{noclone} and the disturbance caused by measurement, could be used to generate secret keys with information-theoretic security. Since then, the field of QKD has grown substantially, giving rise to a wide variety of protocols, including decoy-state QKD~\cite{decoy}, measurement-device-independent QKD~\cite{MDIQKD}, and device-independent QKD~\cite{DIQKD}, all of which aim to enhance practical security under realistic imperfections.

While most experimentally realized QKD schemes rely on photonic encodings transmitted through optical channels~\cite{ph1,ph2,ph3}, there is growing interest in cryptographic primitives that can be implemented directly on gate-based quantum processors. In particular, compatibility with noisy intermediate-scale quantum (NISQ) devices is increasingly important for testing and validating quantum communication protocols at the circuit level~\cite{Kbhar}. Discrete-time quantum walks (DTQWs) provide a promising platform in this direction. A DTQW consists of repeated applications of a coin and a conditional shift operations, leading to coherent spreading and interference across position states~\cite{qw,qw2, DTQWmerits}. Unlike classical random walks, quantum walks exhibit strong interference effects and parameter-dependent dynamical behavior. On cyclic graphs, the nature of the dynamics is highly sensitive to the choice of coin parameters, certain operators generate periodic revivals of the initial state, while others produce chaotic evolution~\cite{Dukes}.

A remarkable phenomenon in discrete‑time quantum walks (DTQWs) is the emergence of Parrondo’s paradox~\cite{Parrondo,Hammer,exptparrando,Abbott}. While originally formulated in classical game theory, its quantum realization has been explored. Two unitary operators that individually generate chaotic dynamics, when combined in a deterministic sequence, can produce fully periodic evolution, a feature that is exact and persistent on finite cycles~\cite{panda,AR}. This interplay between chaos and order offers a natural mechanism for encoding and recovering information. ~\textcolor{black}{Within the specific context of quantum cryptography, Parrondo dynamics have recently attracted attention~\cite{panda,PC1, PC2}. In~\cite{panda, PC1}, it was shown that chaotic switching in quantum coin Parrondo games can generate encrypted keys. Further, ~\cite{PC2} demonstrated that Parrondo-type dynamics can be utilized in a quantum image encryption protocol. In this context, this manuscript focuses on cyclic-graph realizations of Parrondo dynamics and demonstrates their viability as a gate-based cryptographic primitive on NISQ hardware. Our implementation distinguishes itself by not relying on resources such as spatial inhomogeneity or high-dimensional coins, which makes it particularly amenable to circuit-level realization. Accordingly, our proposal constitutes both a secure communication protocol and a QKD-inspired cryptographic primitive that leverages Parrondo dynamics.} Although quantum walk–based cryptographic ideas have been proposed previously \cite{DKP,app1,app2,QKD_Vlachao,Vlachao,r5}, a systematic circuit-level validation under realistic noise models and adversarial conditions remains limited. In particular, it is important to assess how such protocols behave when implemented using gate-based quantum circuits subject to connectivity constraints, state-transfer overhead, and hardware-level noise.

In this work, we present a comprehensive verification of DTQW–based cryptographic protocols that leverage Parrondo’s paradox. Specifically, we design efficient DTQW circuits using QFT-diagonalized shift operators~\cite{r11,AR}, generate public keys from chaotic walk dynamics, encode messages via position-space translation operators, and decrypt messages using the periodic inverse evolution enabled by the Parrondo sequence. To evaluate the usability of such protocols on quantum processors, we employ a module-based simulation framework that assigns separate qubit registers to Alice, Bob, and Eve within a single quantum circuit. Communication steps, which include public-key transfer, message encoding, and adversarial interception, are simulated using SWAP operations. Using both ideal and noise-model simulations in \texttt{qiskit\_aer}~\cite{r14}, we demonstrate accurate message recovery via the Parrondo strategies, robustness of the protocol under realistic quantum noise, and security under intercept–resend and man-in-the-middle attacks~\cite{security}, with Eve’s intervention destroying the walk periodicity required for decryption. \textcolor{black}{A central focus of this manuscript is to bridge the gap between abstract protocol design and practical realization by explicitly incorporating hardware constraints into the construction and analysis of the secure communication protocol.} We analyze the feasibility of implementing the protocol on present-day NISQ hardware, \texttt{ibm\_torino}~\cite{r15}, by mapping logically separated communicating parties onto distant qubit modules and simulating state transfer through SWAP operations and hybrid SWAP-teleportation operations. \textcolor{black}{In particular, we examine how qubit connectivity, state-transfer strategies, and noise accumulation influence the behavior and reliability of the protocol on present-day NISQ devices}. Our results show that protocol performance depends sensitively on qubit choice, transpilation strategy, and hardware topology. 

The manuscript is organized as follows. Sec.~\ref{S2} introduces discrete-time quantum walks on cyclic graphs and the emergence of Parrondo’s paradox. In Sec.~\ref{S3}, we present the Parrondo's paradox-based quantum cryptographic protocol and its gate-level circuit implementation. Sec.~\ref{S4} discusses the numerical results obtained under ideal and noisy simulations. The security of the protocol against intercept-and-resend and man-in-the-middle attacks is analyzed in Sec.~\ref{S5}. Sec.~\ref{S6} provides a comparative discussion with standard quantum cryptographic protocols. The challenges associated with implementing the protocol on present-day NISQ hardware are addressed in Sec.~\ref{S7} and Sec.~\ref{S8} concludes with a summary and outlook. We provide the quantum circuit realizations obtained via Qiskit in Appendix~\ref{A1}. In addition to the circuit realizations in Qiskit, we present proof-of-principle simulations in Mathematica~\cite{Mathematica} in Appendix~\ref{A2}, given the current limitations of NISQ hardware~\cite{Preskill} for distributed implementations~\cite{DQC}. Appendix~\ref{A3} details the NISQ hardware implementation, results, and practical limitations. \textcolor{black}{In Appendix~\ref{A5}, we quantify the security of the proposed cryptographic protocol against any eavesdropping. To further illustrate that the protocol is not restricted to a single Parrondo sequence, we present an additional implementation using a different Parrondo sequence on a 4-cycle graph in Appendix~\ref{A4}.}

\section{Parrondo's Paradox via DTQWs on cyclic graphs} \label{S2}
A DTQW evolving on a $K$-cycle graph is described in the composite Hilbert space, $H_C\otimes H_P$, where $H_P$ denotes the position space and $H_c$ implies the coin space. The coin space $H_C$ is spanned on the basis $\{|l\rangle\}$ i.e., $\{|0\rangle, |1\rangle \} $ and the position space $H_P$ is defined on the vertices of the cyclic graph, \(\{|x\rangle : x \in \{0,1,2,...,K-1\} \} \). Let the walker be initialized at the vertex \( |0\rangle \) and is in a superposition of the  coin states, its initial state is, \begin{equation}
|\Phi(0)\rangle = (\cos\left(\frac{\Theta}{2}\right) |0\rangle + e^{i\omega} \sin\left(\frac{\Theta}{2}\right) |1\rangle) \otimes |K=0\rangle,
\label{e1}
\end{equation} with $0\le \Theta \le \pi$ and $0\le \omega < 2\pi$.
 The coin operator has the form,\begin{equation}
C(s, \gamma, \delta) =
\begin{pmatrix}
\sqrt{s} & \sqrt{1-s} e^{i \gamma} \\
\sqrt{1-s} e^{i \delta} & -\sqrt{s} e^{i(\gamma+\delta)}
\end{pmatrix}, \label{e2}\end{equation} with \( 0 \leq s \leq 1 \) and \( 0 \leq \gamma,\delta \leq \pi \).\par 
The walker moves in counter-clockwise (clockwise) direction by one vertex conditioned on the coin state \( |0\rangle \) (\( |1\rangle \)). The shift operator takes the form \begin{equation} 
    S_K = \sum_{l=0}^{1} |l\rangle \langle l| \otimes |F_l\rangle,
\label{e3} \end{equation} where $F_l = \sum_{x=0}^{K-1} |(x+2l-1) \mod K\rangle \bra{x}$ with $F_0$$(F_1)$ denotes decrement (increment) shift operator corresponding to the counter-clockwise (clockwise) movement of the walker. The unitary single time step evolution operator generating the quantum walk is, \begin{equation}
    W = S_K\cdot (C\otimes I_K), \label{e4}
\end{equation} where $I_K$ is an \( K\times K\) identity operator in the position space and $W$ is a \( 2K\times 2K\)  circulant matrix \textcolor{black}{($\mathcal{C}_K$)}  formulated as~\cite{Dukes},
\begin{align}
W(s,\gamma,\delta) &= \textcolor{black}{\mathcal{C}_K }\Bigg(
\begin{bmatrix} 0 & 0 \\ 0 & 0 \end{bmatrix}_{0},\  
\begin{bmatrix} \sqrt{s} & \sqrt{1-s} e^{i \gamma} \\ 0 & 0 \end{bmatrix}_{1},\  
\cdots \Bigg.  \notag \\ 
& \quad \Bigg. ,\begin{bmatrix} 0 & 0 \\ \sqrt{1-s} e^{i \delta} & -\sqrt{s} e^{i(\gamma + \delta)} \end{bmatrix}_{K-1}
\Bigg).
\label{e5}\end{align} After $t$ steps, the final state is given by $ |\Phi(t) \rangle = W^t
 |\Phi(t=0)\rangle $. If, after $t=T$ steps, i.e., \begin{equation} \ket{\Phi(t=T)} = W^T |\Phi(0)\rangle = |\Phi(0) \rangle,
 \label{e6}\end{equation} then the quantum walk is called ordered (periodic), with period $T$, implying that the walker evolves back to its initial state with probability 1. If \{$|y_j\rangle$\} are the eigenvectors of $W$ and $ \alpha_j $ are the eigenvalues, the initial state $\ket{\Phi(0)}$ can be written as $|\Phi(0)\rangle  = \sum_{i=1}^{2K}a_i|y_i\rangle $ where $a_i$ is the amplitude of the $ith$ eigen vector. After $t$ time steps, we get, \begin{equation} W^t |\Phi(0)\rangle = \sum_{i=1}^{2K}a_i\alpha_i^t|y_i\rangle .\label{e7}\end{equation} Substituting Eq. (\ref{e7}) in Eq. (\ref{e6}) gives, \begin{equation} W^T = I_{2K},  \text{or } \alpha_i^T = 1,\forall\ i \in [1,2K], \label{e8}\end{equation} which defines the criterion for periodic evolution. If $W$ satisfies Eq.(\ref{e8}), it yields a periodic quantum walk. The coin operator $C$ decides whether the quantum walk would be ordered or not. For any arbitrary $K$-cyclic graphs, Ref.~\cite{Dukes} gives the coin parameters that generate periodic walks. \par Periodic quantum walks can also be generated by deterministically combining two chaotic walks, called Parrondo's paradox~\cite{Parrondo,panda,AR}. Here, unitary operators that individually yield chaotic walks with coins, say, $A$ and $B$, are combined in a sequence $AABB..$ to generate a periodic walk for arbitrary cyclic graphs. The chaotic coins $A = C(s = 0.998489, \gamma= 0, \delta = 0 )$ and $B= C(s=0.119545,\gamma=0,\delta=0)$ when combined in a sequence, $AABB...$ produce a periodic walk with period 20 for 4-cycle graphs. Similarly, coins $A' = C(s=0.264734,\gamma=0,\delta=0)$ and $B' = C(s=0.801571,\gamma=0,\delta=0)$ combined in the Parrondo sequence, $A'A'B'B'...$ yield an ordered walk with period 20 in 3-cycle graphs. This interplay between chaos and order provides a mechanism for encoding and recovering information where a chaotic walk can serve as a public state, while the specific Parrondo sequence that restores periodicity remains private.
\section{Parrondo's Paradox-based Quantum cryptographic protocols} \label{S3}

\subsection{Algorithm} \label{S31}
Periodic walks on cyclic graphs generated via DTQW can be exploited to design a quantum cryptographic protocol~\cite{Vlachao}. The protocol uses a public-key generation scheme for a secure encryption-decryption mechanism based on Parrondo's Paradox. Herein, we provide the steps required to execute a secure encryption-decryption of a message on arbitrary cyclic graphs. \\
\textcolor{black}{Before the execution of the protocol, the communicating parties, Alice (receiver) and Bob (sender), are assumed to share a private key ($PvK$) consisting of the set $\{W, G, t, l, x\}$, specifying the time steps required, the initial state of the walker and the associated coin parameters, the evolution operators $W$ and $G$ such that $GW^t = I$ generate the Parrondo sequence, that yield the periodic quantum walk dynamics. This pre-shared information acts as the decryption and authentication key, while the chaotic public-key state is transmitted through the communication channel.}\\
\textbf{1. Generating a chaotic public key.} If Bob intends to send a message $k \in \{0,1,2,...K-1\}$ to Alice, where $k$ is the vertex of the $K$-cyclic graph, then Alice first generates a public key based on the chaotic unitary operator $W$ and the initial state of the walker $\ket{l}\ket{x}$ as, \begin{equation}
    \ket{\Phi_{PK}} = W^t\ket{l}\ket{x}, \label{e33}
\end{equation} where $W$ is the time evolution unitary operator that yields chaotic quantum walks on an $K$-cycle graph, and $\ket{x}$ is the initial position state, while $\ket{l}$ is the initial coin state of the walker. Alice sends the public key, $\ket{\Phi_{PK}}$, to Bob.\\
\textbf{2. Message Encryption.} Bob encodes the message $k$ via,
\begin{equation}
    \ket{\Phi(k)} = (I \otimes T_k)\ket{\Phi_{PK}} ,\label{e34}
\end{equation}  where $I$ is the identity operator defined in the coin space and $T_k = \sum_{i = 0}^{K-1}\ket{(i+k)\text{ mod }K}\bra{i}$ similar to the shift operator described in Eq.~(\ref{e2}) which acts only on the position state, $\ket{x}$ thus, \begin{equation}
    T_k \ket{x} = \ket{(k+x)\text{ mod } K }. \label{e35}
\end{equation} \\
\textbf{3. Message Decryption.} Alice now decrypts the message by applying the Parrondo sequence $G$ on $\ket{\Phi(k)}$ with $GW^t  = I$. Here, it is essential that $[W, T_k] = 0$~\cite{Vlachao}, such that the periodicity condition emerging out of the Parrondo sequence can be satisfied. Further, Alice performs a measurement, $M = \sum_{i=0}^{K-1} I \otimes \ket{i}\bra{i}$ and receives the message $k'$. The original message can be recovered via \begin{equation}
    k = (k' - x) \text{ mod } K .\label{e36}
\end{equation}
\subsection{Quantum Circuit realization} \label{S32}
To realize the cryptographic protocol on the current generation quantum hardware (NISQ devices), we propose a quantum circuit following the approach in Ref.~\cite{AR} for realizing quantum circuits for DTQW on cycles. Herein, we briefly outline the procedure to realize the DTQW on an even ($K = 2^n $) cyclic graph on a quantum circuit. In a gate-based quantum circuit implementation, an 
$K$-cyclic graph is encoded using qubits. For an $K = 2^n$ cyclic graph, the position Hilbert space is represented using 
$n$ qubits, while a single additional qubit encodes the 2D coin space, thus requiring a total of $n+1$ qubits.\\

We illustrate the circuit construction using a 4-cycle graph. The shift operator is, \begin{equation} S_K = \begin{pmatrix} F_0& \mathbf{0}\\ \mathbf{0}& F_1
\end{pmatrix},\label{shift}\end{equation} where $\textbf{0'}$s represent $K\times K$ null matrices in position space, $F_0$ and $F_1$ are the corresponding decrement and increment operators given as,  \begin{equation} F_0 = \begin{pmatrix}
    0& 1& 0& 0 \\
    0& 0& 1& 0 \\
    0& 0& 0& 1 \\
    1& 0& 0& 0 
    \end{pmatrix}, \text{ and}\text{ }F_1 = \begin{pmatrix}
    0& 0& 0& 1 \\
    1& 0& 0& 0 \\
    0& 1& 0& 0 \\
    0& 0& 1& 0 \end{pmatrix}. \label{id} \end{equation} To have an efficient circuit implementation, the shift operators are diagonalized via the quantum Fourier transform (QFT) matrices~\cite{r11}, $M$ and $M^\dagger$ given by, \begin{equation} F_0 = M^\dagger \mathcal{R^\dagger} M \text{ }\text{and}\text{ } F_1 = M^\dagger \mathcal{R} M, \label{dia}\end{equation} where $\mathcal{R(R^\dagger)}$ are the diagonalized shift (decrement and increment) operators with, \begin{equation}
     M = \frac{1}{2}\begin{pmatrix}
            1& 1& 1& 1\\
            1& r& r^2&r^3\\
            1& r^2& r^4&r^6\\
            1& r^3& r^6&r^9\\
            \end{pmatrix}, \label{q}
    \end{equation} with $r = e^{2\pi i/4}$~\cite{Chuang}. Hence, the diagonalized shift operator is given as, \begin{equation} \Delta = (I\otimes M)S_K(I\otimes M^\dagger) = |0\rangle\langle0| \otimes \mathcal{R^\dagger} + |1\rangle\langle1|\otimes \mathcal{R}.\label{e16}\end{equation} QFT is defined on the position space, hence, applies only on the position state of the quantum walker without affecting the coin state. In order to further simplify the implementation of the time evolution operator, one can also express the coin part in terms of the QFT matrices, given as, $(C\otimes I)$ as $(I \otimes M^\dagger)(C\otimes I)(I\otimes M)$.\par Exploiting the property $MM^\dagger = M^\dagger M = I $ results in a quantum circuit that is optimized and efficient by using only one pair of QFT and IQFT operations,  the time evolution operator can thus be written as~\cite{r11}, \begin{equation} W^t = (I \otimes M^\dagger)[\Delta(C\otimes I]^t(I \otimes M). \label{e17}\end{equation} Thus, \begin{equation} \begin{split}
        W^t = (I\otimes M^\dagger)[(|0\rangle\langle0| \otimes I + |1\rangle\langle1|\otimes \mathcal{R}^{2})\\\times(C\otimes \mathcal{R^\dagger})]^t(I\otimes M). \label{e18}\end{split}\end{equation} We employ the above-described method to devise the cryptographic protocol.
\subsubsection{Steps of the protocol}\label{S321}
\begin{figure}
    \centering
    \includegraphics[width=1\linewidth]{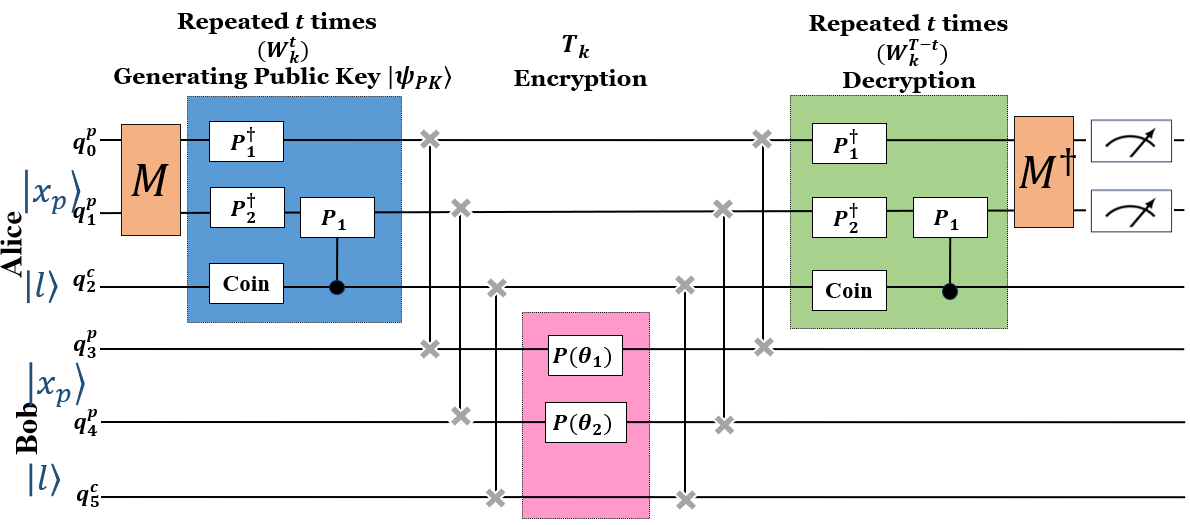}
    \caption{Schematic representation of the quantum circuit of the quantum cryptographic protocol.}
    \label{f1}
\end{figure}

\textbf{1. Generating the chaotic public key. }The initial state of the walker, $\ket{\Phi_i} = \ket{l}\ket{x}$ and the public key generated by Alice from Eq.~(\ref{e33}) is given as, \begin{equation}\begin{split}
    \ket{\Phi_{PK}} = (I \otimes M^\dagger)[(\ket{0}\bra{0} \otimes I + \ket{1}\bra{1} \otimes \mathcal{R}^{2})\\\times(C_B \otimes \mathcal{R^\dagger})]^2(I \otimes M)\ket{\Phi_i} ,\label{e37}
\end{split}\end{equation} where $W^t = BB$ with $C_B$ is the coin operator for unitary $B$. \\
\textbf{2. Public key transfer.} Alice, after generating the public key, sends it to Bob. This is done via the application of SWAP gates between Alice's and Bob's qubits.\\
\textbf{3. Message Encryption. }To encode the message $k$, Bob applies the spatial translation $T_k$ operator (see, Eq. (\ref{e35})) to public key $\ket{\Phi_{PK}}$ with $T_k$ for different values of $k\in \{0,1,2,3\}$ in a $4$-cycle graph, \begin{equation} \begin{split}
     T_0 &= \begin{pmatrix}
        1&0&0&0\\0&1&0&0\\0&0&1&0\\0&0&0&1
    \end{pmatrix}, \quad T_1 = \begin{pmatrix}
        0&0&0&1\\1&0&0&0\\0&1&0&0\\0&0&1&0
    \end{pmatrix},\\ T_2 &=  \begin{pmatrix}
        0&0&1&0\\0&0&0&1\\1&0&0&0\\0&1&0&0
    \end{pmatrix},  \text{ and } T_3 = \begin{pmatrix}
        0&1&0&0\\0&0&1&0\\0&0&0&1\\1&0&0&0
    \end{pmatrix}.\\
\end{split} \label{e38}
   \end{equation} \\ 
   \textbf{4. Message transfer. } After encoding the message, Bob transfers it to Alice, which is done again by the SWAP gates applied to Bob's and Alice's qubits.\\
   \textbf{5. Message decryption. }As Bob encodes the message, Alice applies the decryption operator $G=(AABB)^{4}AA$, since $(AABB)^{5}=I$ (Parrondo sequence with periodicity 20)~\cite{AR}, the final state has the form after two pairs QFTs and IQFTs, one pair while generating the public key and the other pair decrypting the message. The final state is, \begin{equation}
     \begin{split}\ket{\Phi_f} = (I \otimes M^\dagger)[(\ket{0}\bra{0} \otimes I + \ket{1}\bra{1} \otimes \mathcal{R}^{2})(C_G \otimes \mathcal{R^\dagger})]\\\times (I \otimes M)(I \otimes T_k)(I \otimes M^\dagger)[(\ket{0}\bra{0} \otimes I \\+\ket{1}\bra{1} \otimes \mathcal{R}^{2})(C_B \otimes \mathcal{R^\dagger})]^2(I \otimes M)\ket{\Phi_i} ,\end{split}\label{psifinal}\end{equation} $C_G$ being the coin operator associated with unitary $G$. We note that the $T_k$ operators are the same as the shift operators. Hence, they too can be diagonalized by QFT (IQFT) matrices. This lets us use the unitarity condition, $M^\dagger M = I$, and one can have an efficient implementation with a single pair QFT and one IQFT. The diagonalized $T_k$ operator is given as,   \begin{equation}
      T_k^D = M^\dagger T_k M,
    \label{e39}
   \end{equation} where $T_k^D$ is the diagonalized form of $T_k$ which is given as,  \begin{equation}
\begin{aligned}
T_{0}^D &= \begin{pmatrix}
1 & 0 & 0 & 0 \\
0 & 1 & 0 & 0 \\
0 & 0 & 1 & 0 \\
0 & 0 & 0 & 1
\end{pmatrix} = 
\begin{pmatrix}
1 & 0 \\
0 & 1
\end{pmatrix} \otimes 
\begin{pmatrix}
1 & 0 \\
0 & 1
\end{pmatrix}, \\
T_1^D &= \begin{pmatrix}
1 & 0 & 0 & 0 \\
0 & i & 0 & 0 \\
0 & 0 & -1 & 0 \\
0 & 0 & 0 & -i
\end{pmatrix} = 
\begin{pmatrix}
1 & 0 \\
0 & -1
\end{pmatrix} \otimes 
\begin{pmatrix}
1 & 0 \\
0 & i
\end{pmatrix}, \\
T_2^D &= \begin{pmatrix}
1 & 0 & 0 & 0 \\
0 & -1 & 0 & 0 \\
0 & 0 & 1 & 0 \\
0 & 0 & 0 & -1
\end{pmatrix} = 
\begin{pmatrix}
1 & 0 \\
0 & 1
\end{pmatrix} \otimes 
\begin{pmatrix}
1 & 0 \\
0 & -1
\end{pmatrix}, \text{ and } \\
T_3^D &= \begin{pmatrix}
1 & 0 & 0 & 0 \\
0 & -i & 0 & 0 \\
0 & 0 & -1& 0 \\
0 & 0 & 0 & i
\end{pmatrix} = 
\begin{pmatrix}
1 & 0 \\
0 & -1
\end{pmatrix} \otimes 
\begin{pmatrix}
1 & 0 \\
0 & -i
\end{pmatrix}.
\end{aligned}\label{e40}
\end{equation}
Eq.~(\ref{e40}) enables us to encode the message via single-qubit phase rotation gates ($P(\theta)$) i.e., \begin{equation}
    P(\theta) = \begin{pmatrix}
        1&0\\0&e^{i \theta}
    \end{pmatrix}.
\end{equation}
The diagonalized $T_k$'s are thus, \begin{equation} \begin{split}
    T_0 = I \otimes I ,\text{ } T_1 = P(\pi) \otimes P(\pi /2), 
    \text{ }\\T_2 = I  \otimes P(\pi),\text{ and } T_3 =  P(\pi) \otimes P(-\pi /2).
\end{split}
    \label{e46}
\end{equation}
In this work, we propose a module-based quantum cryptography protocol, where communication is modeled through the transfer of quantum states across virtual modules of a quantum processor. Instead of relying on physically separate devices, Alice and Bob are assigned disjoint sets of qubits that act as their respective modules. The exchange of information is simulated using SWAP operations between these registers, enabling the preparation, transmission, and recovery of encoded messages entirely within a single NISQ device. \\
Alice begins with three qubits $q_0,q_1,q_2$ (for $4$-cycle graph), which serve as her position and coin registers. She prepares the initial state by applying coin rotations and conditional shift operators, generating her public key state distributed over these virtual modules. To hand over the state to Bob, Alice uses a sequence of SWAP gates to transfer her prepared qubits into Bob’s register ($q_3,q_4,q_5$). This mimics the physical transmission of a quantum state across a communication channel while remaining entirely within the same quantum processor. Bob encodes his message $k$ through spatial translations (application of $T_k$ operator). After encoding, Bob again uses SWAP gates to send the modified quantum state back to Alice’s registers. With the state returned to her qubits, Alice applies the decryption operator, which is the inverse of her initial public-key operation. Alice then measures her position qubits in the computational basis to read out the message. The schematic representation of the complete protocol is depicted in Fig.~\ref{f1}.

\begin{figure}[t]
\caption{\textbf{Algorithm 1}: Parrondo's Paradox cyclic QW based Cryptographic Protocol on a 4-cycle graph (see, Fig.~\ref{f1})}
\label{a1}
\centering
\fbox{%
\begin{minipage}{0.95\columnwidth}
\small
\begin{algorithmic}[1]
\State \textbf{Input:} Pre-shared private key inputs: Coin operators $A$, $B$ generating chaotic DTQW dynamics, time steps $t$, initial coin $\ket{l}$ and position state $\ket{x}$ of walker 
\State Design the circuit with 6 qubits: Alice 3 qubits $q_0q_1q_2$, Bob 3 qubits $q_3q_4q_5$; Position qubits: $q_0q_1$, $q_3q_4$, Coin qubits: $q_2$ and $q_5$ 
\Statex \textbf{Public Key Generation by Alice}
\State Initialize $q_0q_1q_2$
\State Apply QFT on position qubits of Alice $q_1q_0$ 
\State Apply SWAP on $q_1q_0$ to get the correct output 
\For{$i = 0$ to $2$}
    \State Perform the coin operation B on $q_2$
    \State Perform the phase rotation $P(-\pi)$ on $q_0$  
    \State Perform the phase rotation $P(-\pi / 2)$ on $q_1$
    \If {$\ket{q_2}$ is in $\ket{1}$}
        \State Perform the phase rotation $P(\pi)$ on $q_1$
    \EndIf
\EndFor
\Statex\textbf{Public Key transfer from Alice to Bob}
\State Perform the SWAP gates to swap Alice's qubits with Bob's 
\Statex \textbf{ Message Encryption by Bob}
\State Encrypt the message via $T_k^D$ operator on Bob's position qubits $q_3q_4$
\Statex \textbf{Message transfer from Bob to Alice} 
\State Apply SWAP gates to swap Bob's qubits with Alice 
\Statex \textbf{Message decryption by Alice}
\For{$i = 0$ to $18$} 
    \If{$i \bmod 4 = 0$ or $i \bmod 4 = 1$} 
        \State Perform the coin operation $A$ on $q_2$
    \Else
        \State Perform the coin operation $B$ on $q_2$
    \EndIf
    \State Perform the phase rotation $P(-\pi)$ on $q_0$  
    \State Perform the phase rotation $P(-\pi / 2)$ on $q_1$
     \If {$\ket{q_2}$ is in $\ket{1}$}
        \State Perform the phase rotation $P(\pi)$ on $q_1$
    \EndIf
\EndFor
\State Apply SWAP on $q_1q_0$ to get the correct output 
\State Apply IQFT on position qubits $q_1q_0$
\State Alice measures her position qubits $q_1q_0$
    \end{algorithmic}
    \end{minipage}%
}
\end{figure}

\section{Results} \label{S4}
To implement the proposed cryptographic protocol, we employ the Parrondo strategy in $AABB\ldots$ sequence on a 4-cycle graph, using the parameters for $A$ and $B$ as defined in section II. The quantum circuit, illustrated in Fig.~\ref{f1}, is first simulated using Qiskit’s \texttt{AerSimulator()} to obtain ideal results. Subsequently, noisy simulations are performed with Qiskit’s \texttt{NoiseModel()}.\\ To quantify the similarity between the probability distributions obtained from ideal and noisy discrete-time quantum walk (DTQW) simulations, we employ the \emph{Hellinger fidelity}. Let 
$X = \{x_i\}$ and $Y = \{y_i\}$ denote two discrete probability distributions, where $X$ corresponds to the distribution obtained from ideal simulations (e.g., using \texttt{AerSimulator()}) and $Y$ represents the distribution obtained in the presence of noise. The Hellinger fidelity~\cite{HF,HF_1} between $X$ and $Y$ is formulated as
,\begin{equation}
    H(X,Y) = \left[ 1 - h^2(X,Y) \right]^2,
\label{e22}\end{equation}
with $h$ denoting the Hellinger distance defined as,

\begin{equation}
    h(X, Y) = \frac{1}{\sqrt{2}} 
    \sqrt{\sum_{i} \left(\sqrt{x_i} - \sqrt{y_i}\right)^2}.\label{e23}
\end{equation}

The \textit{Hellinger fidelity} takes values in the interval $[0,1]$, where a value of $1$ indicates identical probability distributions, while $0$ corresponds to completely distinguishable distributions. Fidelity values exceeding $0.5$ signify a strong similarity between the distributions, whereas values above $0.95$ indicate near-perfect agreement. Since our analysis focuses on the dynamics of the probability distributions of the position states, the Hellinger fidelity serves as a natural and meaningful metric to quantify the closeness between the probability distributions acquired from ideal and noisy simulations performed using \texttt{qiskit\_aer}.

\begin{figure}
    \centering
    \includegraphics[width=1\linewidth]{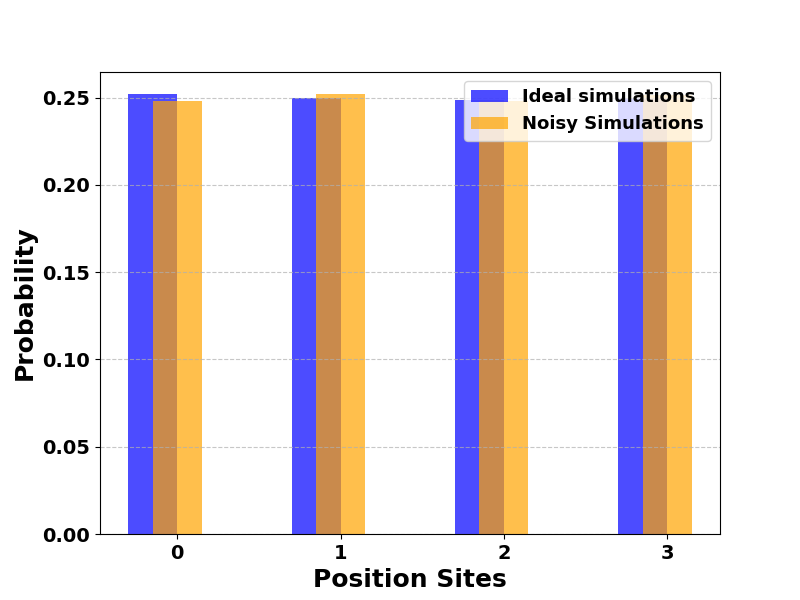}
    \caption{Probability distribution for the public key generation on a 4-cycle graph with initial position $\ket{x} = |0\rangle$ and $\ket{l} = \ket{0}$ implemented in \texttt{qiskit\_aer} with (0.03 depolarizing) and without noise for $10^5$ shots. \textcolor{black}{The close agreement between ideal (blue) and noisy (orange) simulations demonstrates that the public key can be generated reliably under realistic noise conditions.}}
    \label{public}
\end{figure}

\begin{figure}
    \centering
    \begin{subfigure}{0.65\linewidth}
        
        \includegraphics[width=\linewidth]{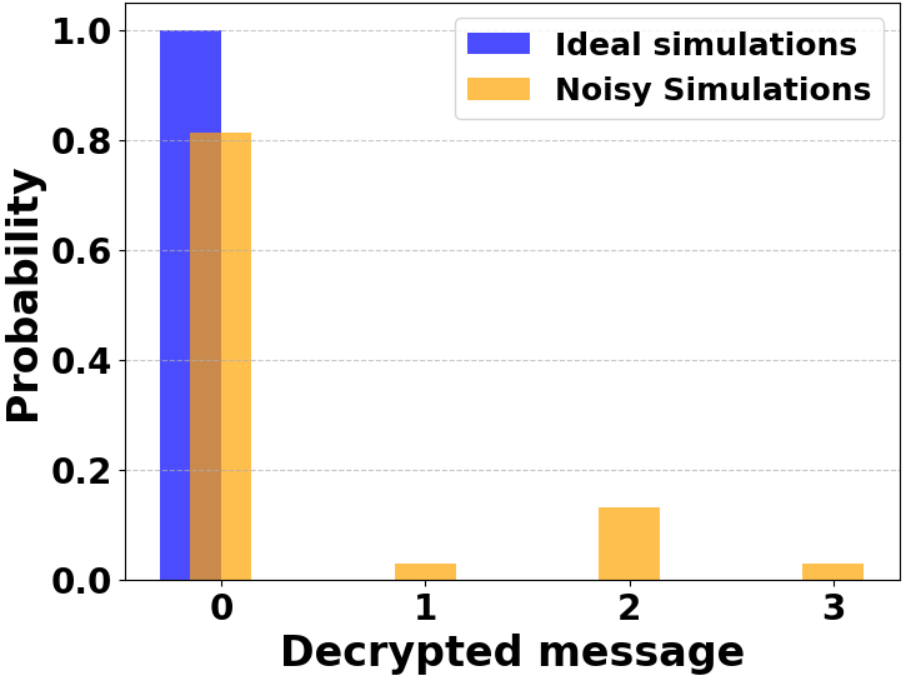}
        \caption{}
    \end{subfigure} %
    \begin{subfigure}{0.65\linewidth}
        
        \includegraphics[width=\linewidth]{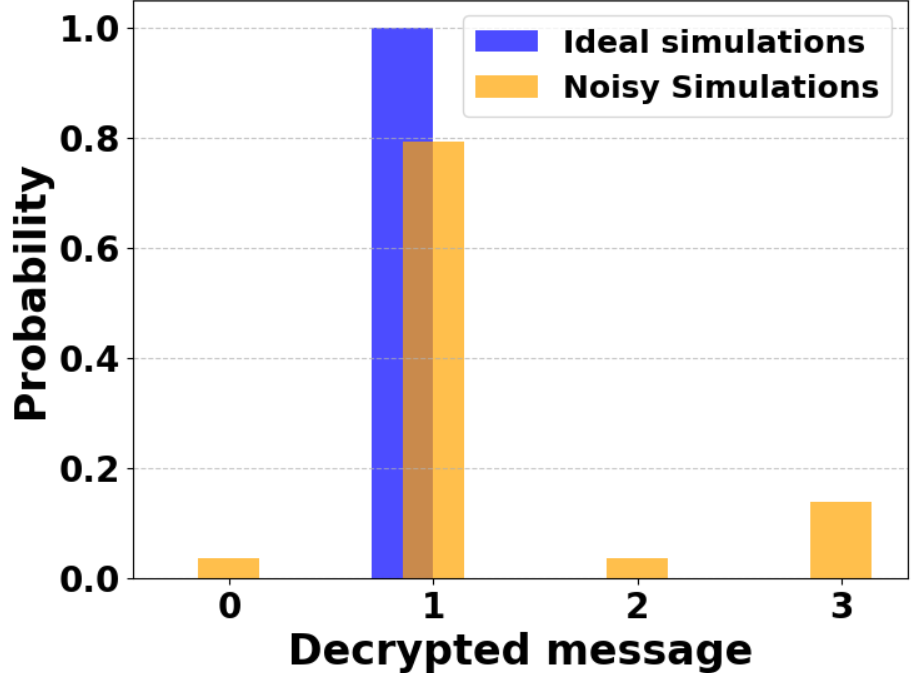}

        \caption{}
    \end{subfigure}
    \begin{subfigure}{0.65\linewidth}
        
        \includegraphics[width=\linewidth]{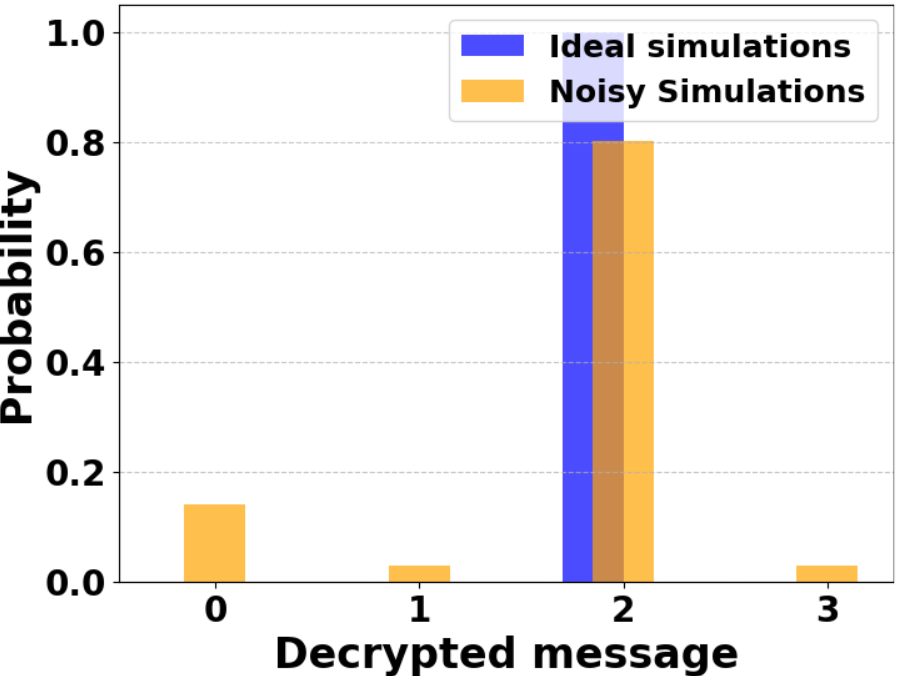}

        \caption{}
    \end{subfigure}
    \begin{subfigure}{0.65\linewidth}
        
        \includegraphics[width=\linewidth]{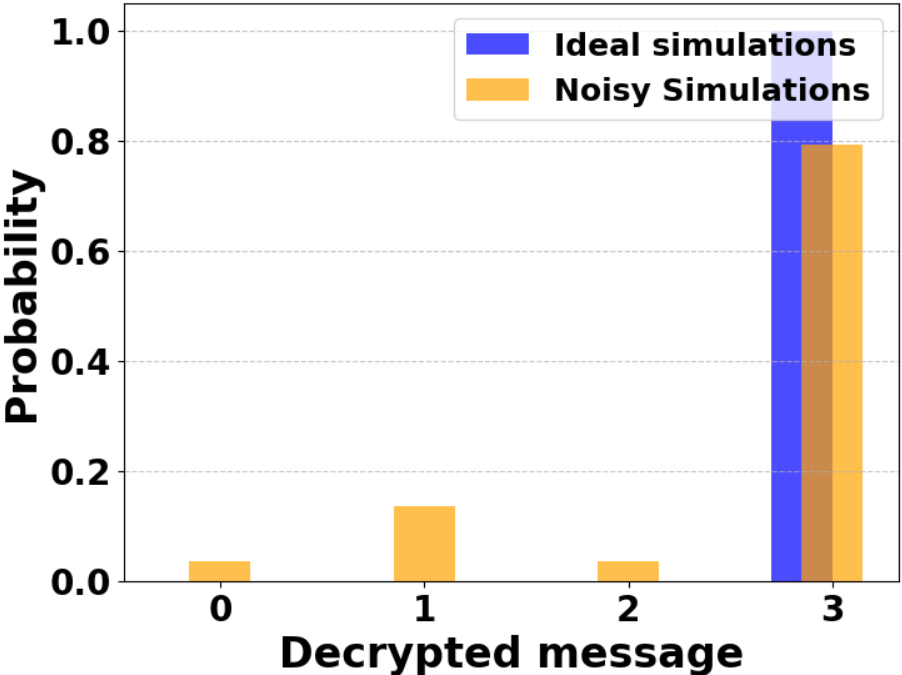}

        \caption{}
    \end{subfigure}
    \caption{Probability distribution for the Decrypted message, $k'$ for encoded message (a) $k = 0 $, (b) $k = 1 $, (c)$k = 2 $, (d) $k = 3 $ with initial position $\ket{x} = |0\rangle$ such that $k'=k$ implemented in \texttt{qiskit\_aer} with depolarizing noise and without noise for $10^5$ shots on a 4-cycle graph. \textcolor{black}{The peak at the correct message state in both ideal and noisy simulations demonstrates successful recovery of the encoded message, indicating that the periodic Parrondo dynamics required for decryption remain robust under realistic noise conditions.}}
    \label{f2}
\end{figure}

\begin{figure}[H]
    \centering
    \includegraphics[width=1\linewidth]{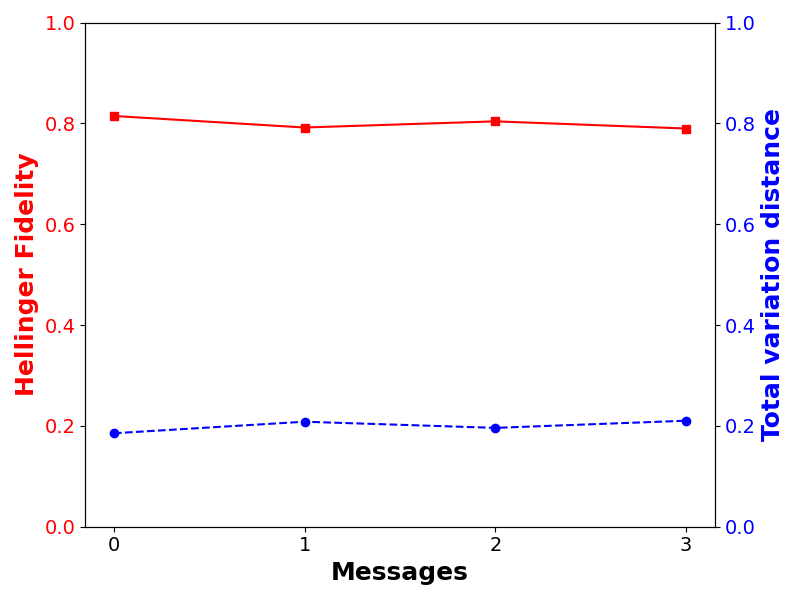}
    \caption{Hellinger Fidelity and total variance distance for different messages decrypted by Alice with initial position $\ket{x} = \ket{0}$ and coin $\ket{s} = \ket{0}$ implemented in \texttt{qiskit\_aer} with depolarizing noise and without noise for $10^5$ shots on a 4-cycle graph.}
    \label{HF}
\end{figure}

\begin{figure}
\centering
    \begin{subfigure}{0.65\linewidth}
        \centering
        \includegraphics[width=\linewidth]{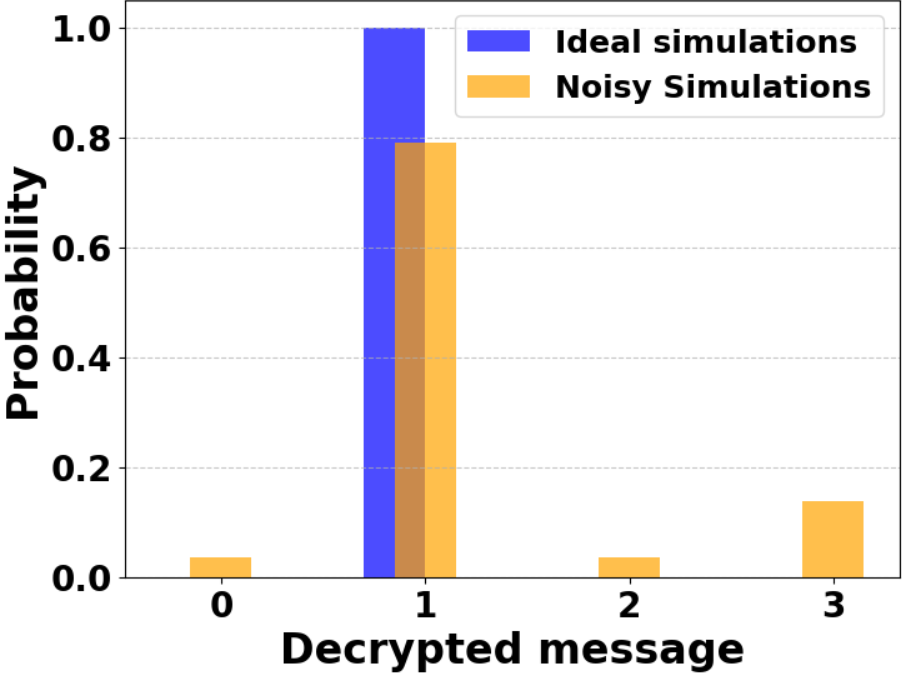}
        \caption{}
    \end{subfigure} %
    \begin{subfigure}{0.65\linewidth}
        \centering
        \includegraphics[width=\linewidth]{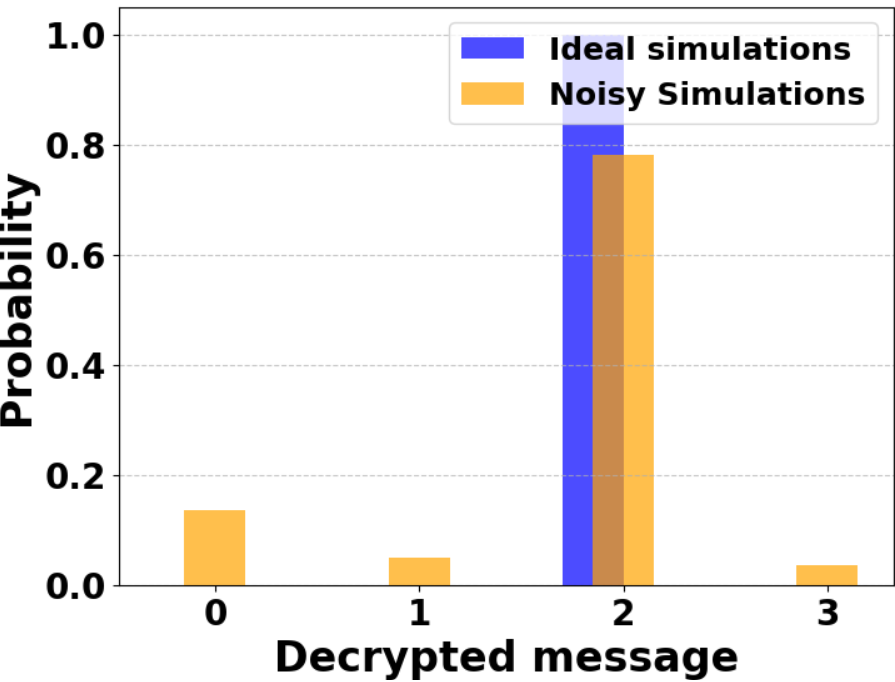}

        \caption{}
    \end{subfigure}
    \begin{subfigure}{0.65\linewidth}
        \centering
        \includegraphics[width=\linewidth]{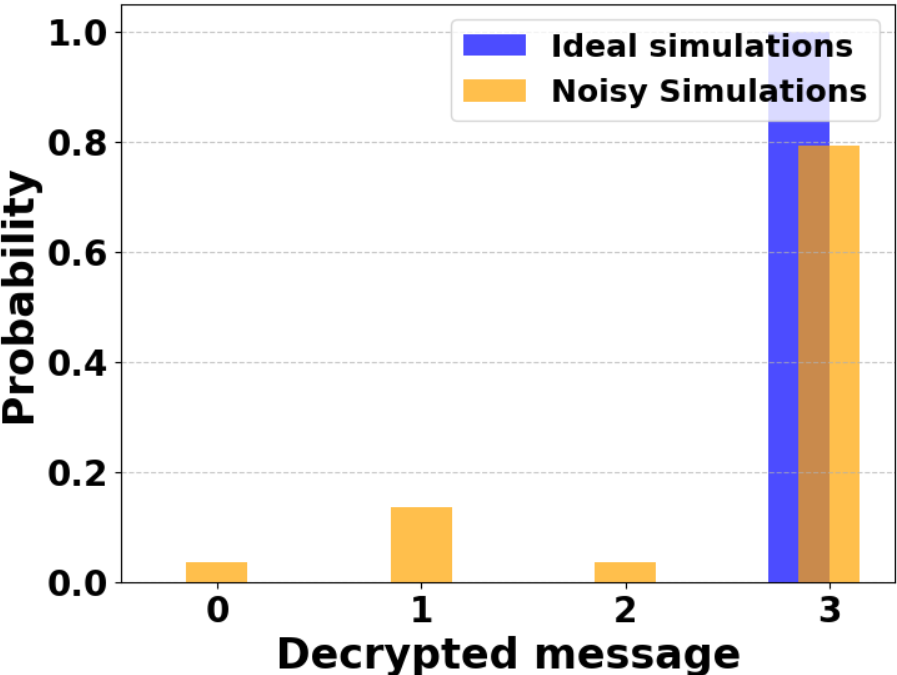}

        \caption{}
    \end{subfigure}
    \begin{subfigure}{0.65\linewidth}
        \centering
        \includegraphics[width=\linewidth]{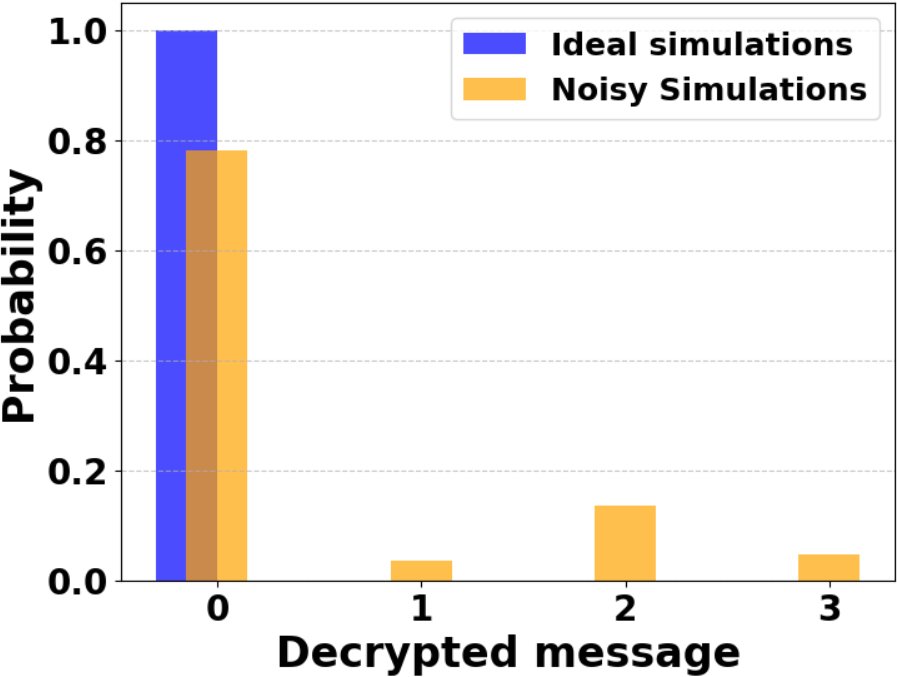}

        \caption{}
    \end{subfigure}
    \caption{Probability distribution for the Decrypted message, $k'$ for encoded message $k = 1$ for different initial position (a)$\ket{x} = |0\rangle$, (b) $\ket{x} = |1\rangle$, (c) $\ket{x} = |2\rangle$, and (d) $\ket{x} = |3\rangle$ such that $k'= (k+x) \text{ mod} \text{ K}$, with $l \in \{0,1,2,3\}$ implemented in \texttt{qiskit\_aer} with depolarizing noise and without noise for $10^5$ shots on a 4-cycle graph. \textcolor{black}{The peak at the correct message state in both ideal and noisy simulations demonstrates successful recovery of the encoded message irrespective of the initial state, indicating that the periodic Parrondo dynamics required for decryption remain robust under realistic noise conditions.}}
    \label{f4}
      \end{figure}

In addition to the Hellinger fidelity, one may also evaluate the total variation distance, which measures the statistical distinguishability between two probability distributions by quantifying the maximum difference in the probabilities that each distribution assigns to the same events. For probability distributions  \( X = \{x_i\} \) and \( Y = \{y_i\} \), the total variation distance is given by
\begin{equation}
    T (X,Y) = \frac{1}{2} \sum_{k} |x_i - y_i|. 
\end{equation}
The total variation distance is bounded by $[0,1]$, where 0 indicates the distributions are similar, while 1 means that they are completely different. 
\\Fig. \ref{public} shows the probability distribution during the public key generation of the cryptographic protocol, with the walker initialized at $\ket{0}$ position state and $\ket{l} = \ket{0}$ coin state. \textcolor{black} {It illustrates the public key generation stage, where the quantum walk evolves into a near uniform position-space distribution. While this may appear visually flat, it encodes an important physical feature of the protocol, which is the formation of a chaotic key.} The results show the circuit realizations both in ideal conditions and with realistic quantum noise using Qiskit’s noise modeling framework. Gate imperfections are modeled by a depolarizing quantum channel with an error probability of 0.03 applied uniformly to all quantum gates in the circuit. The depolarizing channel provides an effective description of averaged gate errors arising from control inaccuracies, crosstalk, and residual decoherence, and is commonly used to benchmark quantum algorithms on NISQ devices. The chosen error probability closely reflects the effective noise levels observed on current-generation IBM superconducting quantum processors~\cite{r15} operated at optimization level 3. This approach allows us to capture the aggregate influence of gate noise on the quantum walk dynamics and the resulting cryptographic performance. The close similarity between the ideal and noisy probability distributions indicates that the protocol is stable and that noise has only a small effect on the quantum walk dynamics. This means that the public key can be generated accurately even on the current generation noisy intermediate-scale quantum (NISQ) hardware. For completeness, the quantum circuits implementing the protocol are presented in Appendix~\ref{A1}. In addition to the Qiskit-based simulations presented above, we also provide a proof-of-principle implementation of the complete cryptographic protocol using the \texttt{Wolfram Quantum Framework}~\cite{WQF} in Mathematica. The corresponding results of the ideal and noisy simulations are reported in Appendix~\ref{A2}.\\
Fig.~\ref{f2} presents the decrypted messages obtained when the quantum circuits are realized with and without noise, with nearly 80\% Hellinger fidelity (see, Fig.~\ref{HF}) for all the messages. Additionally, Fig. \ref{f4} shows that the messages encoded using different initial walker states can also be reliably decrypted, with Hellinger fidelity, 80\%, and total variation distance, 20\%, further demonstrating the robustness and correctness of our circuit design. The slightly reduced fidelity observed in the noisy simulations can be understood as a consequence of how the protocol is realized at the circuit level. The encryption–decryption process involves several sequential operations, including repeated SWAP gates used to transfer quantum states between the virtual modules representing Alice and Bob. These additional operations increase the overall circuit depth, allowing noise to accumulate over the course of the computation. On real quantum hardware, such state transfers typically correspond to larger physical separations between qubits and limited connectivity, which further amplify gate errors. As a result, the final probability distribution becomes more spread out. Importantly, the fidelity remaining close to 80\% indicates that the underlying periodic dynamics of the quantum walk is largely preserved, enabling reliable message recovery despite realistic noise levels. \textcolor{black}{This behavior highlights that the observed performance is not solely determined by the underlying protocol, but is strongly shaped by circuit-level implementation constraints, reinforcing the importance of hardware-aware design in evaluating quantum cryptographic schemes.}


\section{Security of the protocol} \label{S5}
\subsection{Intercept and Resend Attack} \label{S51}
In quantum communication protocols, the security of information exchange between two legitimate parties, Alice and Bob, relies on the fundamental principles of quantum mechanics. However, an adversary, commonly referred to as Eve, may attempt to gain unauthorized access to the transmitted quantum states. One of the most basic yet impactful strategies employed by Eve is the intercept-and-resend attack~\cite{security}.

In this attack model, Eve intercepts the quantum states sent from Alice to Bob, performs measurements on them, and then prepares and sends new quantum states to Bob that mimic the original as closely as possible. By doing so, Eve attempts to extract information about the secret data or key being exchanged, while minimizing the disturbance introduced into the system. Since quantum states cannot, in general, be measured without altering them, Eve’s intervention inevitably introduces detectable errors in the communication channel. The appearance of such anomalies alerts Alice and Bob to the presence of eavesdropping.\\
On the security of our cryptographic protocol, let us consider the DTQW on a 4-cycle graph such that the position state be $\ket{x} = |00\rangle$ (i.e., site $\ket{K=0}$), and coin state $|l\rangle = |0\rangle$ initially, and the message is, say, $ k=1$. An eavesdropper can attack at step 3 (stages 12-13 see, Algorithm 1 (Fig.~\ref{a1}) when Bob sends the encrypted
 and encoded message $\ket{\Phi(k)}$ to Alice. Since Eve does not know the private key, Eve learning the state $|\Phi(k = 1)\rangle$ is almost impossible (i.e., it has negligible probability). This mitigates
 an intercept-and-resend attack wherein Eve can try to measure Alice’s signal or the message-encrypted state.
To assess the security of the protocol at the circuit level, we introduce an eavesdropper, Eve, equipped with three qubits analogous to the modules used by Alice and Bob. Eve is then incorporated into the communication process, enabling us to simulate an intercept-and-resend attempt within the same computational framework. By executing the protocol in the presence of Eve and tracking the resulting state evolution, we can evaluate how effectively the cryptographic scheme detects and resists unauthorized interference. The complete algorithm demonstrating the protocol’s security under this attack model is presented in Algorithm 2 (Fig.~\ref{a2}). 

\begin{figure}
    \centering
    \includegraphics[width=1\linewidth,scale=0.7]{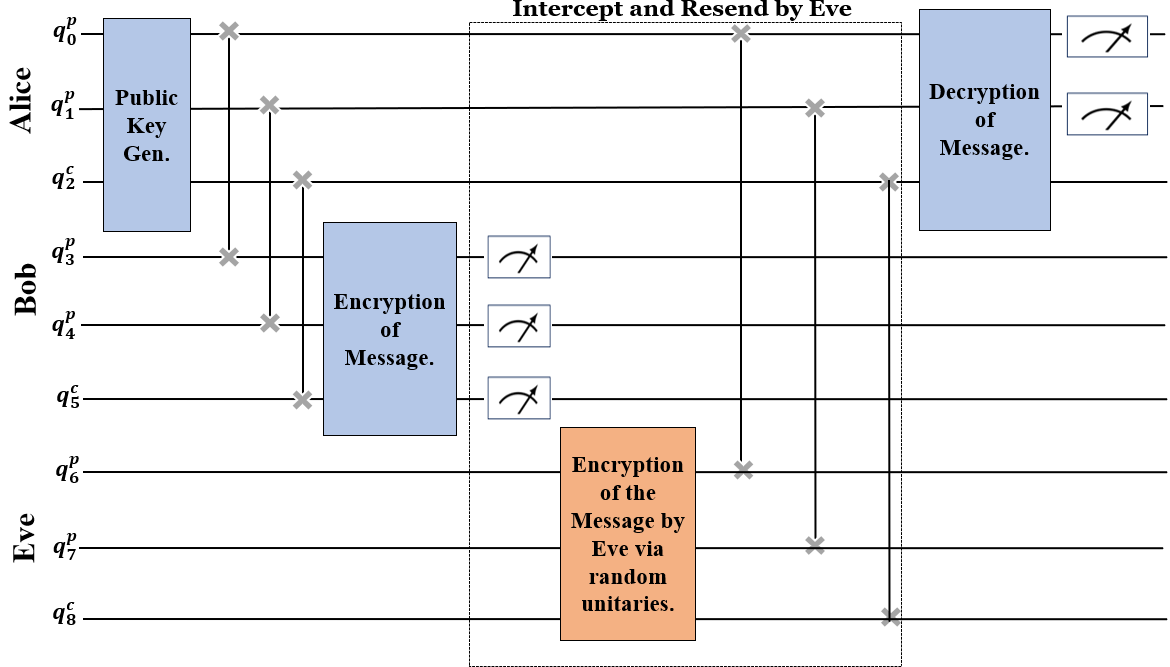}
    \caption{Schematic representation of the quantum circuit in the presence of Eve performing an intercept and resend attack on the quantum cryptographic protocol.}
    \label{Eve_ckt_scheme}
\end{figure}

To quantify the effect of Eve’s intercept-and-resend attack, we evaluate the Quantum Bit Error Rate (QBER), defined here as the probability that Alice’s decrypted message differs from the correct value. For our DTQW-based protocol, QBER~\cite{qber} is computed as, \begin{equation}
    \text{QBER} = 1-P(k),
    \end{equation} where $P(k)$ is the probability (from the final position-state distribution) that Alice recovers the correct message. QBER is bounded in the interval [0,1], where a value of 0 indicates error-free message recovery and a value of 1 corresponds to complete failure of decryption. Exact decryption in the proposed protocol relies on the periodic dynamics of the quantum walk under the private Parrondo sequence. In the presence of Eve, the collapse of the walk’s periodic structure spreads the probability across all position states, sharply decreasing the probability of getting the correct message and yielding a high QBER. Thus, QBER serves as a clear indicator of Eve’s interference in intercept-and-resend attacks. A schematic representation of the quantum circuit with Eve performing an intercept and resend attack is depicted in Fig.~\ref{Eve_ckt_scheme}.

\twocolumngrid
\begin{figure}[t]
\caption{\textbf{Algorithm 2}: Intercept and Resend Attack on 4-cycle graphs (see, Fig.~\ref{Eve_ckt_scheme})}
\label{a2}
\centering
\fbox{%
\begin{minipage}{0.95\columnwidth}
\small
\begin{algorithmic}[1]
\State \textbf{Input:} Pre-shared private key inputs: Coin operators $A$, $B$ generating chaotic DTQW dynamics, time steps $t$, initial coin $\ket{l}$ and position state $\ket{x}$ of walker
\State Design the circuit with 9 qubits: Alice 3 qubits $q_0q_1q_2$, Bob 3 qubits $q_3q_4q_5$, and Eve 3 qubits $q_6q_7q_8$. Position qubits for Alice are $q_0q_1$, for Bob are $q_3q_4$, and for Eve are $q_6q_7$. Coin qubits: Alice $q_2$, Bob $q_5$, and Eve $q_8$
\Statex \textbf{Public Key Generation by Alice}
\State Initialize $q_0q_1q_2$
\State Apply QFT on position qubits of Alice $q_1q_0$ 
\State Apply SWAP on $q_1q_0$ to get the correct output 
\For{$i = 0$ to $2$}
    \State Perform the coin operation B on $q_2$
    \State Perform the phase rotation $P(-\pi)$ on $q_0$  
    \State Perform the phase rotation $P(-\pi / 2)$ on $q_1$
    \If{$\ket{q_2}$ is in $\ket{1}$}
        \State Perform the phase rotation $P(\pi)$ on $q_1$
    \EndIf
\EndFor
\Statex\textbf{Public Key transfer from Alice to Bob}
\State Apply SWAP gates to swap Alice's qubits with Bob's 
\Statex \textbf{ Message Encryption by Bob}
\State Encrypt the message via the $T_k^D$ operator to Bob's position qubits $q_3q_4$
\Statex \textbf{Interception by Eve}
\State Eve measures Bob's qubits by applying random (computational or X basis) measurement operators 
\State Eve encodes the message based on her measurements via random unitary matrices on qubits $q_6q_7q_8$
\Statex \textbf {Message transfer by Eve}
\State Apply SWAP gates to swap Eve's qubits with Alice's 
\Statex \textbf{Message decryption by Alice}
\For{$i = 0$ to $18$} 
    \If{$i \bmod 4 = 0$ or $i \bmod 4 = 1$} 
         \State Perform the coin operation $A$ on $q_2$
    \Else
        \State Perform the coin operation $B$ on $q_2$
    \EndIf
    \State Perform the phase rotation $P(-\pi)$ on $q_0$  
    \State Perform the phase rotation $P(-\pi / 2)$ on $q_1$
    \If{$\ket{q_2}$ is in $\ket{1}$}
        \State Perform the phase rotation $P(\pi)$ on $q_1$
    \EndIf
\EndFor
\State Apply SWAP on $q_1q_0$ to get the correct output 
\State Apply IQFT on position qubits $q_1q_0$
\State Alice measures her position qubits $q_1q_0$
    \end{algorithmic}
    \end{minipage}%
}
\end{figure}

\begin{figure}
\centering
    \begin{subfigure}{0.7\linewidth}
        
        \includegraphics[width=\linewidth]{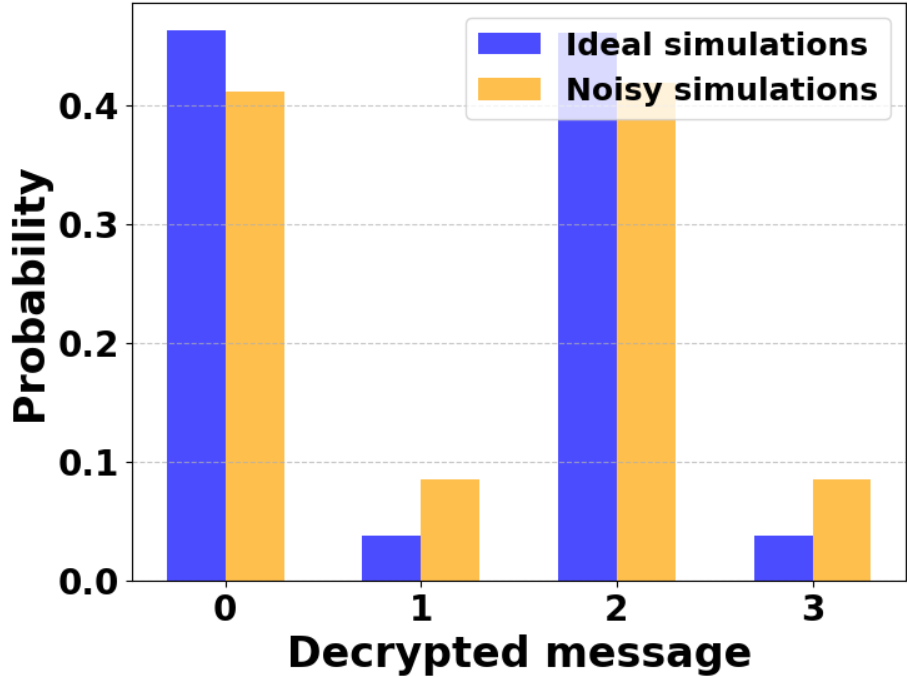}
        \caption{}
    \end{subfigure} %
    \begin{subfigure}{0.75\linewidth}
        
        \includegraphics[width=\linewidth]{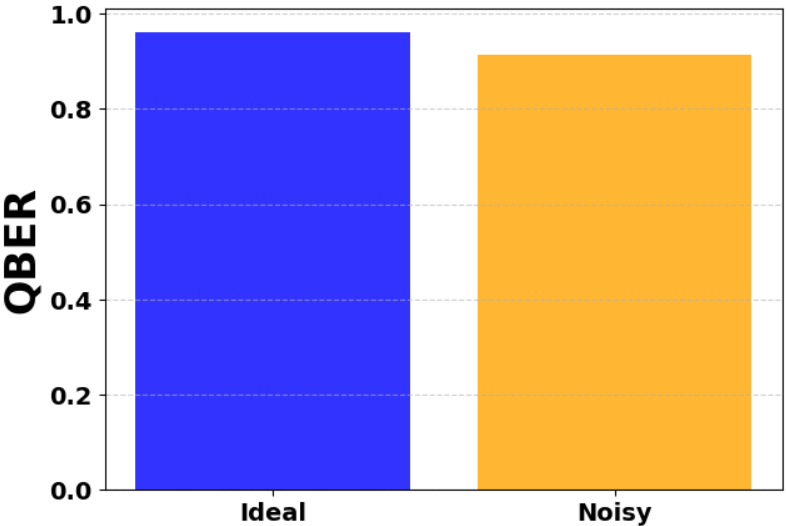}
        \caption{}
    \end{subfigure}
    \caption{(a) Probability Distribution for the Decrypted message  $k'$, intercepted by Eve for encoded message $\text{ } k = 1 $ with initial position $\ket{x} = |0\rangle$ such that $k'=k$ implemented in \texttt{qiskit\_aer} with depolarizing noise and without noise for $10^5$ shots. (b) Quantum bit error rate (QBER) obtained in ideal and noisy simulations for the cryptographic protocol intercepted by Eve. \textcolor{black}{The broad distribution of outcomes and the maximal QBER indicate that Eve's measurement destroys quantum-walk interference, leading to decryption failure and providing a clear signature of eavesdropping.}}
\label{f5}
\end{figure}%
Fig.~\ref{f5} (a) presents the probability distribution of the decrypted message obtained by Alice with Eve performing an intercept and resend attack on the encoded message $k = 1$. The results from both ideal and noisy simulations show that measurement outcomes are widely dispersed across all possible decrypted messages rather than concentrated around the correct value. Instead of recovering the message $k'= 1$ (see Fig.~\ref{f2}), the presence of Eve causes Alice’s subsequent decryption to yield a mixed distribution over all message values, indicating that the original message information is effectively lost as a result of Eve’s interference. This mixture ensures that Eve's attempt not only fails to reveal the encoded message but also introduces detectable disturbances in the communication process. The noisy simulation further broadens this distribution, reinforcing the conclusion that Eve gains no advantage even in non-ideal environments. \textcolor{black}{Overall, this confirms that the protocol demonstrates strong robustness against the intercept-and-resend attack considered in this study.} 
Eve’s intrusion transforms the decrypted outcome into a mixture of all possible messages, ensuring both message confidentiality and attack detectability. Also, Eve’s measurement causes the probability at the correct outcome to drop to a very small value (ideal--0.04 and noisy -- 0.08) while the remaining probability mass spreads uniformly across the incorrect messages. Consequently, the QBER rises close to its maximal value (ideal--0.96 and noisy--0.92), indicating that Alice almost never recovers the correct message when Eve intervenes. This extremely high QBER (see Fig.~\ref{f5} (b))confirms that any intercept-and-resend attempt produces a strong, unambiguous disturbance in the decrypted outcomes, making the attack immediately detectable. The corresponding circuit realizations used in the security analysis and \textcolor{black}{detailed discussion on the security proof are included in Appendices~\ref{A1} and \ref{A5}}.

\subsection{Man-in-the-middle Attack} \label{S52}
The man-in-the-middle attack~\cite{security} can be viewed in this context as an authentication check, where Alice verifies whether the received quantum state originates from Bob or has been altered by an adversary such as Eve. Bob and Alice share a pre-established private key in the form of a Parrondo sequence, which ensures that the quantum walk exhibits periodic dynamics during decryption. \textcolor{black}{If an adversary intercepts, modifies, or replaces the transmitted state, the applied operations generally disrupt this structured evolution. Consequently, when Alice performs the Parrondo sequence, the resulting position distribution deviates from the expected periodic pattern, providing a clear signature of tampering.\\}
\begin{figure}[t]
\caption{\textbf{Algorithm 3}: Authentication check in man-in-the-middle attack performed by Eve}
\label{am3}
\centering
\fbox{%
\begin{minipage}{0.95\columnwidth}
\small
\begin{algorithmic}[1]
\State \textbf{Input:} Coin operators $A$, $B$ generating chaotic DTQW dynamics, time steps $t$, initial coin $\ket{l}$ and position state $\ket{x}$ of walker
\State Design the circuit with 6 qubits: Alice 3 qubits $q_0q_1q_2$, Bob 3 qubits $q_3q_4q_5$. Position qubits for Alice are $q_0q_1$, and for Bob are $q_3q_4$. Coin qubits: Alice $q_2$, and Bob $q_5$.
\Statex \textbf{Public Key Generation by Alice}
\State Initialize $q_0q_1q_2$
\State Apply QFT on position qubits of Alice $q_1q_0$ 
\State Apply SWAP on $q_1q_0$ to get the correct output 
\For{$i = 0$ to $2$}
    \State Perform the coin operation B on $q_2$
    \State Perform the phase rotation $P(-\pi)$ on $q_0$  
    \State Perform the phase rotation $P(-\pi / 2)$ on $q_1$
    \If{$\ket{q_2}$ is in $\ket{1}$}
        \State Perform the phase rotation $P(\pi)$ on $q_1$
    \EndIf
\EndFor
\Statex\textbf{Public Key transfer from Alice to Bob}
\State Apply SWAP gates to swap Alice's qubits with Bob's 
\Statex \textbf{ Message Encryption by Bob}
\State Encrypt the message via the $T_k^D$ operator to Bob's position qubits $q_3q_4$

\Statex \textbf {Message transfer by Bob}
\State Apply SWAP gates to swap Bob's qubits with Alice's 
\Statex \textbf{Man-in-the-middle attack and decryption of message by Eve}
\For{$i = 0$ to $10$} 
    \State Eve applies random unitaries to decrypt the message
    \EndFor
\State Apply SWAP on $q_1q_0$ to get the correct output 
\State Apply IQFT on position qubits $q_1q_0$
\State Measure the position qubits $q_1q_0$
    \end{algorithmic}
    \end{minipage}%
}\end{figure}

\begin{figure}
\centering
    \begin{subfigure}{0.7\linewidth}
        
        \includegraphics[width=\linewidth]{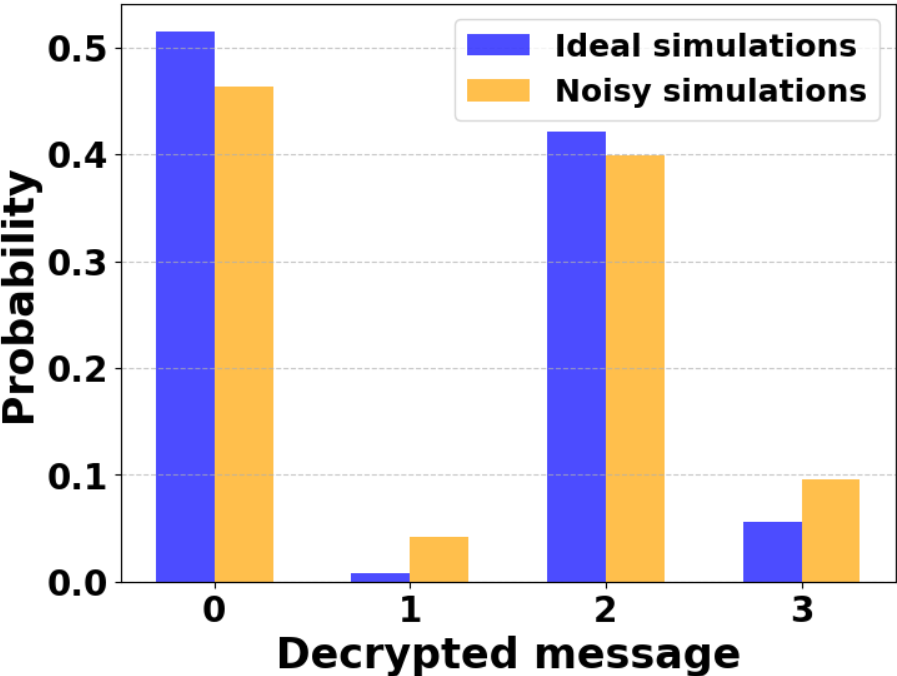}
        \caption{}
    \end{subfigure} %
    \begin{subfigure}{0.75\linewidth}
        
        \includegraphics[width=\linewidth]{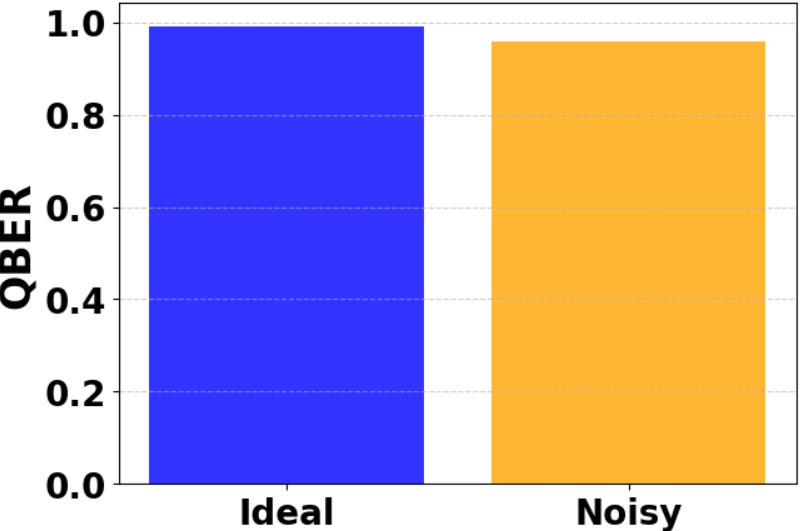}
        \caption{}
    \end{subfigure}
    \caption{\textcolor{black}{(a) Probability Distribution for the Decrypted message  $k'$, when decrypted by Eve for encoded message $\text{ } k = 1 $ with initial position $\ket{x} = |0\rangle$ such that $k'=k$ implemented in \texttt{qiskit\_aer} with depolarizing noise and without noise for $10^5$ shots. (b) Quantum bit error rate (QBER) obtained in ideal and noisy simulations for the cryptographic protocol decrypted by Eve. High QBER indicates that Eve's arbitrary decryption attempts fail to reproduce the Parrondo dynamics, preventing correct message recovery and revealing unauthorized intervention.}}
\label{mitm}
\end{figure}%
\textcolor{black}{An adversary attempting a man-in-the-middle attack must reconstruct both the underlying Parrondo sequence and the associated coin parameters governing the quantum walk. The difficulty arises from the sensitivity of the walk dynamics to these parameters; small deviations in the inferred coin angles lead to noticeable discrepancies in the resulting probability distributions. As a result, even approximate reconstruction produces measurable errors, which can be quantified using metrics such as the quantum bit error rate (QBER). The algorithm demonstrating the authentication check under this attack model is presented in Algorithm 3 (Fig.~\ref{am3}). The circuit-level realization of man-in-the-middle attacks is included in Appendix~\ref{A1}.} Furthermore, the chaotic dynamics in the public key ($\ket{\Phi_{PK}}$) add another layer of protection by producing a highly sensitive, interference-dominated quantum state that is effectively impossible to invert or replicate without knowledge of the exact Parrondo sequence.
\textcolor{black}{This behavior is illustrated in Fig.~\ref{mitm}, where Eve attempts to decrypt the encrypted state without knowledge of the exact private Parrondo sequence (decryption operator $G$) shared between Alice and Bob. Instead of the correct deterministic sequence, Eve applies random unitary operations for arbitrary time steps (Step 18 of Algorithm 3), which fail to recover the interference structure required for periodic reconstruction of the quantum walk. Consequently, the decrypted probability distribution spreads across all possible message states rather than peaking at the correct message ($k = 1$), and the QBER remains high in both ideal ($\approx 99\%$) and noisy ($\approx 95\%$) simulations. This demonstrates that unauthorized decryption attempts using arbitrary unitaries cannot reproduce the correct Parrondo dynamics required for message recovery.}

\section{Analysis} \label{S6}
The results presented in the previous section demonstrate that our DTQW-based cryptographic protocol performs reliably under ideal and noisy conditions, and exhibits a very strong disturbance signature under intercept-and-resend attacks. To appreciate the significance of this behavior from a broader quantum-cryptographic perspective, it is useful to relate our findings to what is typically expected from a standard prepare-and-measure protocol. BB84 is the most widely recognized example of such a scheme, and serves as a conventional benchmark for how noise and eavesdropping reveal themselves through quantum state disturbance. \textcolor{black}{We use BB84 here purely as a familiar reference point to contextualize the disturbance signature of our protocol under a simulated intercept-and-resend attack. Since BB84 and the present protocol operate on fundamentally different principles and serve different cryptographic functions, the comparison is intended to be qualitative rather than a direct assessment of relative performance.}
\begin{table}[h!]
\centering
    \renewcommand{\arraystretch}{1.1}

\begin{tabular}{|l|c|c|c|}
\hline
\textbf{Scenario} & \textbf{BB84~\cite{bb84}} & \textbf{Parrondo DTQW} \\
\hline
Ideal (no Eve) & $\approx 0$ & $\approx 0$ \\
\hline
Eve intercept--resend & $\approx 0.25$ & $0.92$ \\
\hline
Authentication & \shortstack{Needs extra \\classical channel} & \shortstack{Inherent \\(Private key)} \\
\hline

\end{tabular}
\caption{Comparison between BB84 and the proposed Parrondo's paradox-based DTQW protocol based on QBER values}
\label{t1}
\end{table}
\par Table~\ref{t1} compares the performance of BB84 protocol and our proposed Parrondo's paradox-based DTQW protocol obtained from simulations of both schemes under (i) ideal conditions, and (ii) a full intercept-and-resend attack. To establish a baseline, we simulated the BB84 protocol using a single-qubit prepare-and-measure scheme in Qiskit. In the ideal case, the sifted keys obtained by Alice and Bob match perfectly, resulting in a QBER $\approx0$. The sifted key is formed after Alice and Bob publicly compare their measurement bases and retain only those bits for which compatible bases were used, discarding the rest. Under an intercept-and-resend attack, Eve collapses the prepared quantum state by measuring it in a randomly chosen basis and resending her outcome. This induces a well-known disturbance of approximately 0.25 in the sifted key. Our implementation reproduces this behavior, confirming that BB84 detects eavesdropping statistically through a characteristic but relatively mild increase in QBER. This detection requires a sufficiently large number of transmitted qubits for reliable estimation. The results of our proposed protocol demonstrate a qualitatively different behavior. The periodic evolution reliably recovers the encrypted messages with high success probability. The most notable effect appears when Eve attempts an intercept-and-resend attack. By performing a measurement, Eve collapses the multi-amplitude quantum-walk state and is unable to reproduce the private coin dynamics. Consequently, Alice’s decryption no longer yields peaked distributions; instead, it approaches a near-uniform output. The probability of recovering the correct message falls to approximately 0.04–0.08 (see, Fig.~\ref{f5} (a)), corresponding to QBER values in the range 0.92–0.96 (see, Fig.~\ref{f5} (b)), \textcolor{black}{which is notably larger than the characteristic disturbance of 0.25 induced in BB84 under the same attack model, reflecting the different ways in which the two protocols respond to adversarial measurement.}
Thus, any adversarial intervention results in almost total failure of decryption, enabling immediate and unmistakable detection without the need for statistical sampling. In addition, the BB84 protocol does not have native authentication. To authenticate the legitimate communicating parties, an additional classical channel is required, whereas our proposed protocol ensures an \textcolor{black}{implicit authentication feature, as Eve’s intervention disrupts the periodic reconstruction,} enabling secure message transfer and instantaneous eavesdropper detection from a single run rather than relying on statistical averaging over long keys. \textcolor{black}{This work is primarily a demonstration of a secure quantum communication protocol based on Parrondo dynamics, implemented at the circuit level on NISQ-compatible architectures. At the same time, the protocol is inspired by QKD in the sense that it leverages disturbance under measurement as a signature of adversarial intervention. We emphasize that the work is not intended as a replacement for BB84‑style QKD protocols, but rather as a demonstration of how quantum walk dynamics can serve as a new cryptographic resource. More broadly, the framework can be viewed as a cryptographic primitive, where information encoding, transmission, and recovery are governed by structured quantum walk dynamics. }
\section{NISQ implementation: Results and Challenges}~\label{S7}
The proposed Parrondo's paradox-based cryptographic protocol was implemented and tested on IBM’s superconducting quantum computing platform using the \texttt{ibm\_torino} backend, a 133-qubit quantum device based on the \textit{Heron r1} processor, accessed via IBM Quantum cloud services~\cite{r15}. Realizing the proposed protocol on present-day NISQ hardware presents several practical challenges that arise from limited qubit connectivity, hardware noise, and transpiler-induced circuit transformations~\cite{Preskill}. In principle, the protocol assumes that Alice and Bob reside on spatially separated quantum processors and exchange quantum states over a communication channel. Since such distributed quantum computation is not yet reliably supported on current NISQ hardware~\cite{DQC}, we emulated this setting on a single processor by assigning disjoint and widely separated qubit modules to Alice and Bob. In our implementation, Alice and Bob were mapped to distant registers, thereby maximizing logical separation as detailed in Appendix~\ref{A3}.\\
A primary challenge stems from quantum state transfer between distant qubit modules. Quantum state transfer between Alice and Bob was realized via sequences of SWAP operations across intermediate qubits. This leads to a substantial increase in circuit depth and, consequently, significant error accumulation. Moreover, at higher transpiler optimization levels, aggressive qubit remapping and gate reordering can partially or entirely erase the intended modular separation between Alice and Bob, resulting in a loss of distinction between the communicating parties and undermining the cryptographic structure of the protocol (see, Appendix~\ref{C1}). To mitigate these limitations, we explored hybrid quantum state-transfer strategies combining SWAP operations with quantum teleportation, which more closely resemble a distributed communication paradigm while preserving logical modularity. As shown in Appendix~\ref{C2}, these hybrid approaches can better maintain the Alice–Bob separation enforced at the circuit level. However, teleportation introduces its own sources of error through Bell-pair generation, additional entangling gates, and measurement operations. Consequently, the relative performance of SWAP-then-teleport versus teleport-then-SWAP strategies is found to depend sensitively on the specific choice of physical qubits, their connectivity, and local error rates.\\Overall, these results demonstrate that the protocol performance is governed by a trade-off between circuit depth, noise accumulation, and preservation of logical communication structure. The findings underscore the necessity of hardware-aware qubit selection and communication ordering when implementing cryptographic protocols on current quantum processors and motivate our simulation-based, proof-of-principle validation. Although this limitation prevented a \textcolor{black}{demonstration} on distributed hardware, recent advances toward modular quantum architectures~\cite{rec} represent promising steps toward fully distributed implementations in the future.

\section{Conclusion} \label{S8}
Periodic quantum walks can be used to design a quantum cryptographic protocol, where Parrondo sequences combine individually chaotic operators to obscure the walk dynamics while still allowing periodicity to be recovered with the correct sequence. Consequently, the encryption stage generates a highly delocalized quantum state, and successful decryption depends on applying the precise unitary sequence that restores periodicity. Building on this idea, we propose an explicit quantum circuit realization of DTQW–based cryptographic protocol that is compatible with the current generation gate-based NISQ hardware~\cite{r15}. Given the current limitations of NISQ hardware, which do not yet support implementations across physically separated quantum processors, we realize the protocol on a single quantum processor by assigning disjoint qubit modules to represent the communicating parties, with Alice as the receiver and Bob being the sender. We analyze the performance of the protocol through numerical simulations carried out in Qiskit under both ideal conditions and in the presence of depolarizing noise. In the noiseless case, the periodic structure of the walk is faithfully recovered, leading to high agreement between the decrypted and original probability distributions. When depolarizing noise is included, the recovered distributions deviate from the ideal case, which is quantitatively reflected by Hellinger fidelity ($\approx 80\%$) and the total variation distance ($\approx 20\%$). These deviations arise from accumulated gate errors and state-transfer operations inherent to the circuit implementation. The security of the protocol arises from encoding the message onto a chaotic public key generated by the DTQW. In the circuit implementation, this public key corresponds to a chaotic unitary evolution, making the encoded state highly sensitive to interference. Any action by an eavesdropper alters the implemented gate sequence and leads to observable deviations in the decrypted statistics. In contrast, successful authentication requires the application of the exact decryption operator, implemented as a specific Parrondo sequence of coin operations. \textcolor{black}{The private Parrondo sequence acts simultaneously as a decryption key and as a structurally inherent authenticator; successful message recovery certifies possession of the correct private sequence, and any adversarial interference manifests directly as decryption failure, within the attack models considered in this study}. By explicitly modeling intercept-and-resend and man-in-the-middle attacks within the quantum circuit framework, we showed that adversarial intervention disrupts the periodic dynamics, leading to a pronounced increase in QBER ($\approx 92\%$) and a clear statistical signature of interference. These results indicate that the protocol exhibits strong disturbance sensitivity at the circuit level. We further corroborate our findings through independent simulations in the Wolfram Quantum Framework (Mathematica), ensuring platform-independent validation of the protocol dynamics (detailed in Appendix~\ref{A2}). \textcolor{black}{A key outcome of this study is the demonstration that the feasibility and performance of such protocols are intrinsically linked to hardware-level considerations.} As detailed in Appendix~\ref{A3}, state transfer realized solely through SWAP operations introduces significant limitations. At higher transpiler optimization levels, the logical distinction between communicating parties is compromised, whereas at lower optimization levels the increased circuit depth leads to reduced fidelity. To address these constraints, we investigate hybrid state-transfer strategies that combine SWAP operations with quantum teleportation. While this approach improves overall protocol performance, it remains sensitive to the choice of physical qubits and the underlying qubit connectivity of the device. 
\\While this manuscript focuses on small cyclic graphs, the framework can be extended to larger systems and more realistic hardware-specific error channels. Exploring the implementation on physically distributed quantum processors, optimization of state-transfer strategies, and analysis under more general adversarial models mark an important direction for future study. More broadly, this work demonstrates that dynamical phenomena intrinsic to quantum walks, rather than entanglement or basis randomness alone, can serve as the foundation for secure, circuit-level quantum cryptographic protocols in near-term quantum devices.
\onecolumngrid
\appendix
\section{Quantum circuit realizations in \texttt{QISKIT}} \label{A1}
\begin{figure}[H]
    \centering
    \includegraphics[width=0.7\linewidth]{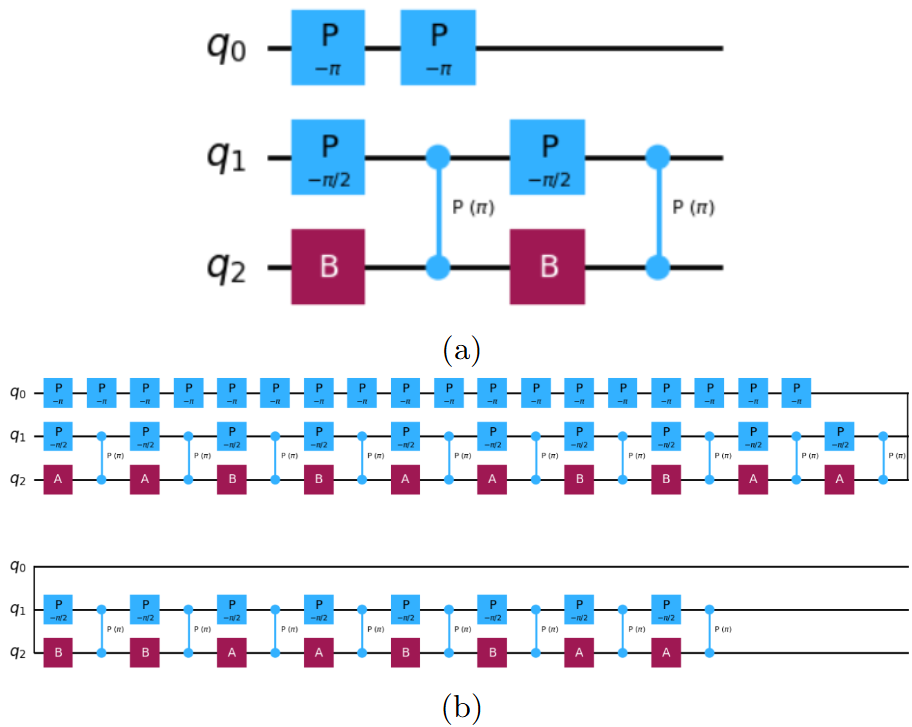}
    \caption{Quantum circuits for (a) Public Key generated by Alice ($\ket{q_oq_1q_2}$) and (b) Decryption operator $G = (AABB)^{4}AA$ applied by Alice to recover the message encrypted by Bob.}
    \label{Pk}
\end{figure}

\begin{figure}[H]
    \centering
    \includegraphics[width=0.95\linewidth]{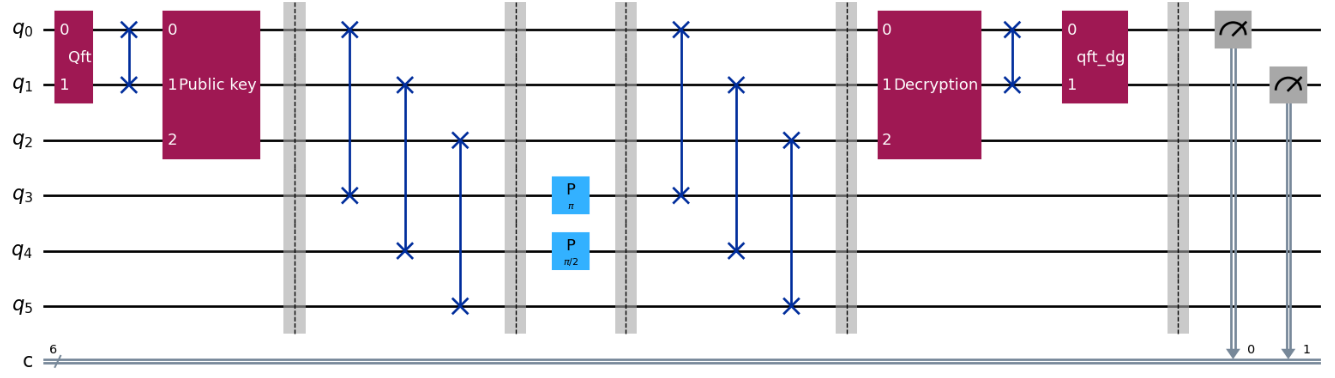}
    \caption{Quantum circuit implementing the proposed Parrondo’s paradox–based cryptographic protocol. Qubits $q_0,q_1$ and $q_2$ represent Alice’s position and coin degrees of freedom, while qubits $q_3,q_4$ and $q_5$ correspond to Bob’s register. The circuit realizes public key generation by Alice with initial state $|000\rangle$, message $k = 1$ encoding by Bob, state transfer between the parties, and subsequent decryption performed by Alice.}
    \label{totalckt}
    \end{figure}

Herein, we present the explicit quantum circuits used to implement and verify the Parrondo’s paradox–based cryptographic protocol within the Qiskit framework. The circuits shown here correspond to the building blocks and full protocol discussed in Sec.~\ref{S32} and are provided to ensure transparency and reproducibility of the results reported in Secs.~\ref{S4} and~\ref{S5}.
All circuits are constructed using a gate-based model suitable for near-term NISQ devices. For a 4-cycle graph, the DTQW is encoded using three qubits: two qubits represent the position space, while one qubit denotes the coin degree of freedom. As described earlier (see, Sec.~\ref{S7}), we implement the protocol with Alice and Bob as separate modules in a single quantum processor. In the module-based cryptographic implementation, Alice and Bob are each assigned independent three-qubit registers, and communication between them is modeled using SWAP operations within a single quantum processor.\\Fig.~\ref{Pk} illustrates the two fundamental circuit components used in the cryptographic protocol. Fig. 9(a) shows the public key generation step implemented by Alice, where the initial walker state is evolved through successive applications of the chaotic coin operator $B$, following the DTQW dynamics described in Sec.~\ref{S3}. This operation generates the public key that is subsequently shared with Bob. Fig. 9(b) depicts the decryption operator $G$ employed by Alice to recover Bob’s encrypted message. The operator $G$ corresponds to the deterministic Parrondo sequence $(AABB)^{4}AA$ implemented as a sequence of DTQW steps.\\
Fig.~\ref{totalckt} shows the complete quantum circuit implementing the proposed cryptographic protocol. The circuit incorporates public key generation, message encoding by Bob, state transfer between Alice and Bob, and the subsequent decryption performed by Alice. 

\begin{figure}[H]
    \centering
    \includegraphics[width=1\linewidth]{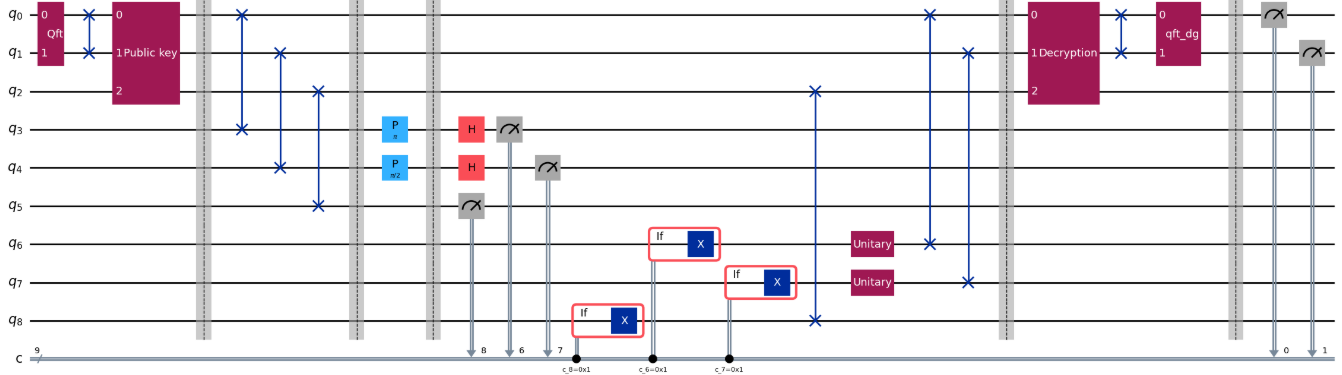}
    \caption{Quantum circuit implementing the proposed Parrondo’s paradox–based cryptographic protocol in the presence of an adversary Eve. Qubits $q_0,q_1$ and $q_2$ represent Alice’s position and coin degrees of freedom, while qubits $q_3,q_4$ and $q_5$ correspond to Bob’s register, and qubits $q_6,q_7$ and $q_8$ correspond to Eve.}
    \label{Eveckt}
\end{figure}

\begin{figure}[H]
    \centering
    \includegraphics[width=1\linewidth]{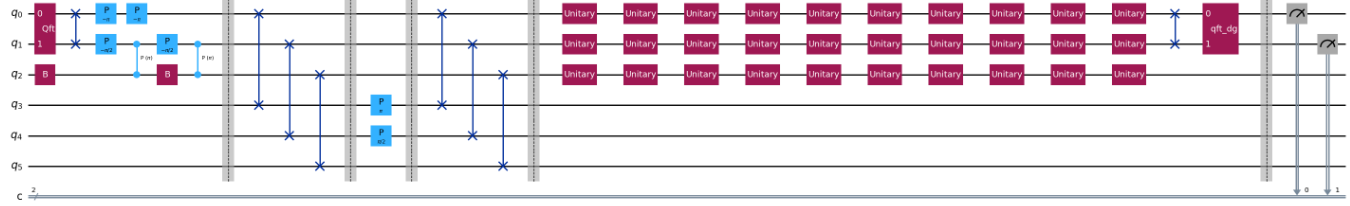}
    \caption{\textcolor{black}{Quantum circuit implementing the authentication check in the proposed Parrondo’s paradox–based cryptographic protocol in the presence of an adversary Eve. Eve attempts to decrypt the intercepted encrypted state by applying random unitary operations instead of the correct private Parrondo sequence.}}
    \label{Mitmckt}
\end{figure}

Fig.~\ref{Eveckt} presents the extended circuit used to simulate adversarial behavior, where an additional three-qubit register is assigned to Eve. This circuit realizes an intercept-and-resend attack by allowing Eve to measure, re-encode, and forward the quantum state to Alice. \textcolor{black}{Fig.~\ref{Mitmckt} shows the authentication-check circuit in the presence of Eve, where the adversary attempts to decrypt the intercepted encrypted state using random unitary operations instead of the correct private Parrondo sequence.} The resulting disruption of the walk’s periodicity is directly observable in the decrypted probability distributions and forms the basis of the security analysis discussed in Sec.~\ref{S5}.
All circuits were simulated using Qiskit Aer under both ideal conditions and with realistic depolarizing noise models, as described in Sec.~\ref{S4}. The appendix figures thus serve as a circuit-level reference for the numerical simulations and security analysis presented in the main text. The codes supporting the plots of the main text are available in Ref.~\cite{github}.
\section{Quantum Circuit Realization and Simulation in Mathematica} \label{A2}
In addition to the Qiskit-based implementation presented in the main text, we independently realized and simulated the Parrondo’s paradox–based quantum cryptographic protocol using the \texttt{Wolfram Quantum Framework}~\cite{WQF} in Mathematica. \textcolor{black}{It is important to note that the noise models employed in the two frameworks are not identical and should not be interpreted as directly equivalent.} \\In the Qiskit framework, realistic noise is modeled by applying noise channels to individual unitary gates, closely mimicking hardware-level gate imperfections. In contrast, the Mathematica simulations employ an effective qubit-level depolarizing channel, \textcolor{black}{applied to the final state, which captures decoherence and operational imperfections}. Specifically, noise is modeled using a depolarizing quantum channel, with two different cases of depolarizing probabilities, 0.03 and 0.2~\cite{WQF_1}. The depolarizing channel provides an effective description of decoherence and operational imperfections acting directly on the qubit state. The lower noise level, 0.03, is chosen to enable a direct and fair comparison with the gate-level noise strength employed in the Qiskit simulations. The higher noise level, 0.2, is introduced to probe the robustness of the protocol under stronger decoherence and to examine how the fidelity degrades as noise increases. This allows us to assess the sensitivity of the protocol to noise strength and to gain insight into the range of noise models relevant for more general and large-scale studies.
The depolarizing channel is applied after the full encryption–decryption circuit, acting independently on all six qubits (two position qubits and one coin qubit for both Alice and Bob). This approach captures the cumulative effect of noise on the final quantum state, rather than attributing errors to specific gates, and is suited for studying the robustness of the protocol against global qubit decoherence. 

\begin{figure}[h]
    \centering
    \includegraphics[width=1\linewidth]{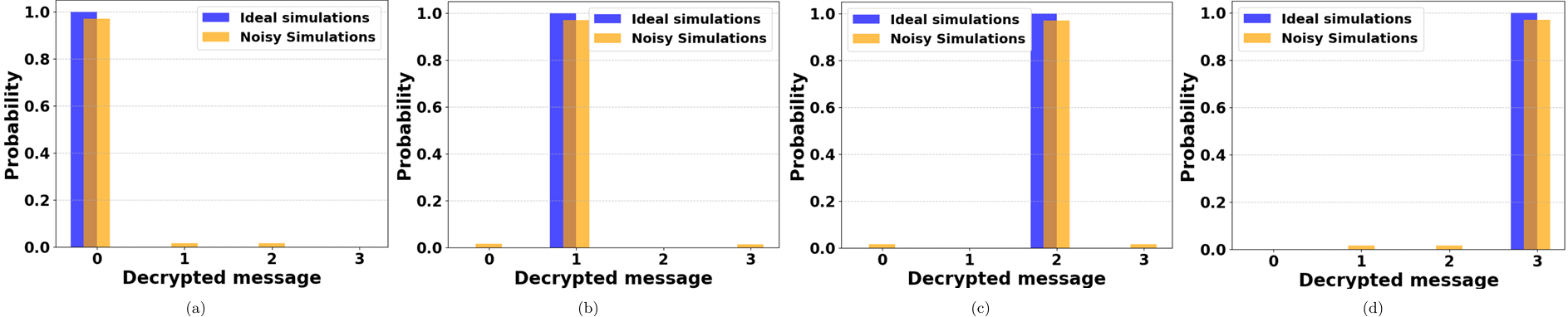}
    \caption{Probability distribution for the Decrypted message, for encoded message $(a)\text{ } k = 0 $,$ (b)\text{ } k = 1 $,$ (c)\text{ } k = 2 $,$(d)\text{ } k = 3 $ with initial position of the walker $\ket{x} = |0\rangle$ such that the decrypted message is same as the encoded message ($k'=k$) implemented in \texttt{Wolfram Quantum Framework} with depolarizing noise (3\%) and without noise for $10^5$ shots on a 4-cycle graph.}
    \label{WQF}
\end{figure}
\begin{figure}[H]
    \centering
    \includegraphics[width=1\linewidth]{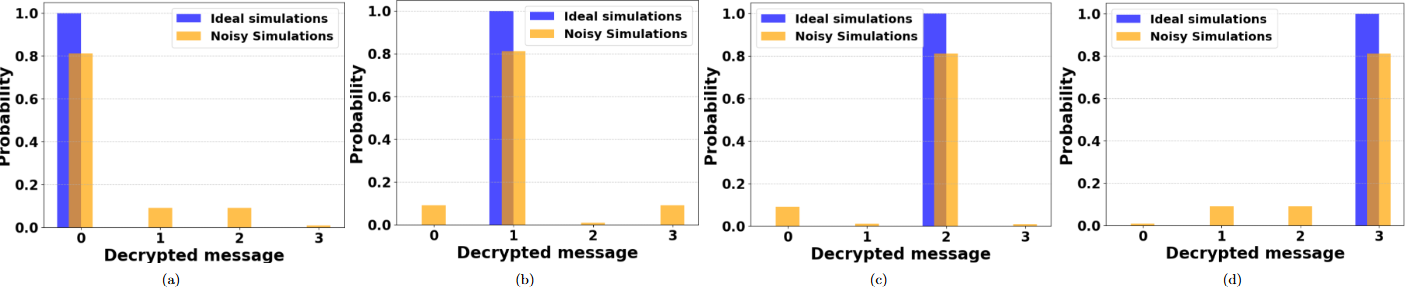}
    \caption{Probability distribution for the Decrypted message, for encoded message $(a)\text{ } k = 0 $,$ (b)\text{ } k = 1 $,$ (c)\text{ } k = 2 $,$(d)\text{ } k = 3 $ with initial position of the walker $\ket{x} = |0\rangle$ such that the decrypted message is same as the encoded message ($k'=k$) implemented in \texttt{Wolfram Quantum Framework} with depolarizing noise (20\%) and without noise for $10^5$ shots on a 4-cycle graph.}
    \label{WQF_more}
\end{figure}
\begin{figure}[h]
    \centering
    \includegraphics[width=0.7\linewidth]{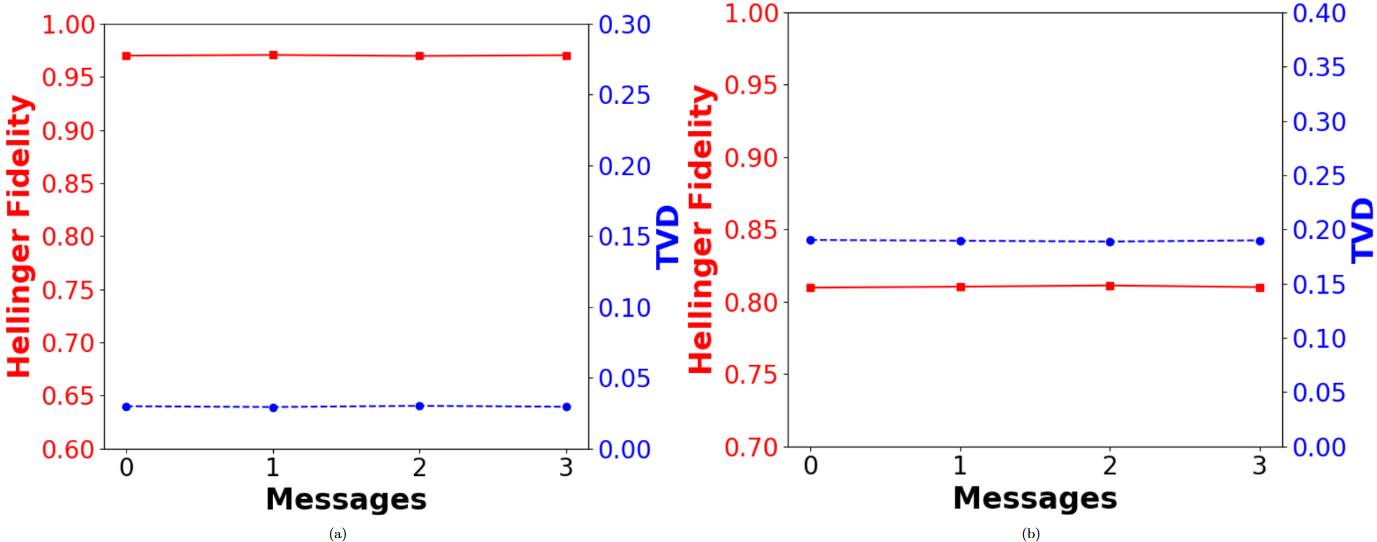}
    \caption{Hellinger Fidelity and total variance distance for different messages decrypted by Alice with initial position $\ket{x} = \ket{0}$ and coin $\ket{l} = \ket{0}$ implemented in \texttt{Wolfram Quantum Framework} with (a) less noise (3\%) and (b) more noise (20\%) for $10^5$ shots on a 4-cycle graph.}
    \label{WQF_HF}
\end{figure}
In the absence of noise, the decrypted probability distributions obtained from Mathematica exhibit a sharply peaked structure at the correct message value, reflecting the periodic dynamics induced by the Parrondo sequence. For each encoded message $k\in$ {0,1,2,3}, the probability of recovering the correct decrypted message $k' = k$ (for initial state $|x\rangle = |0\rangle$) approaches unity, while the probabilities of the remaining position states are strongly suppressed. This behavior is consistent with the ideal Qiskit simulations and confirms that the Parrondo evolution accurately retraces the public-key dynamics. When depolarizing noise is introduced at the qubit level, the probability distributions broaden, as shown in Figs.~\ref{WQF} and \ref{WQF_more} for depolarizing probabilities 0.03 and 0.2. Despite this broadening, the correct decrypted message remains the most probable outcome in all cases. In particular, for the weak noise case, 0.03, the success fidelity of the correct decrypted message is $\approx 97\%$ (see Fig.~\ref{WQF_HF} (a)), indicating near-ideal recovery despite the presence of realistic qubit-level decoherence. Even for the stronger noise strength, 0.2, the dominant peak corresponding to the correct message persists with fidelity $\approx81\%$ (see Fig.~\ref{WQF_HF} (b)), indicating that the periodic dynamics survives substantial decoherence. \textcolor{black}{The near-constant values indicate message-independent performance. Increasing noise reduces fidelity and increases variation distance, quantifying the protocol’s degradation.}
 Quantitatively, the noisy Mathematica results show a redistribution of probability mass from the correct message to neighboring position states, while preserving a clear statistical bias toward the correct outcome. This behavior mirrors the trends observed in the Qiskit simulations, where noise reduces fidelity but does not eliminate successful decryption. \textcolor{black}{Comparisons between the two frameworks are intended to be qualitative rather than quantitative. Nevertheless, we observe that a depolarizing strength of approximately 20\% in the qubit-level Mathematica model produces a comparable level of degradation in the output probability distributions as a 3\% gate-level depolarizing noise in the Qiskit simulations.} The agreement between the two platforms confirms that the observed robustness is an intrinsic feature of the Parrondo-driven discrete-time quantum walk dynamics. The comparative performance of the protocol implemented in \texttt{qiskit\_aer} and the \texttt{Wolfram Quantum Framework}, under varying noise strengths, is summarized in Table~\ref{t2}. The Mathematica-based realization of the protocol and related simulation framework are presented in Ref.~\cite{Wpage}.

\begin{table}[h!]
\centering
    \renewcommand{\arraystretch}{1.1}
\begin{tabular}{|l|c|c|c|c|}
\hline
\textbf{Methods} & \textbf{Noise type} &\textbf{Noise level }&\textbf{\shortstack{Hellinger \\Fidelity}} &\textbf{\shortstack{Total Variation\\distance}} \\
\hline
\texttt{qiskit\_aer} & Depolarizing Gate noise &0.03 &$\approx 80\%$ &$\approx 20\%$ \\
\hline
\texttt{Wolfram Quantum Framework} & Depolarizing qubit noise& 0.03 &$\approx 97\%$ & $\approx 3\%$ \\
\hline
\texttt{Wolfram Quantum Framework} & Depolarizing qubit noise& 0.2 &$\approx 81\%$ & $\approx 19\%$ \\
\hline

\end{tabular}
\caption{Comparison between the results obtained for the proposed cryptographic protocol implemented in \texttt{qiskit\_aer} and \texttt{Wolfram Quantum Framework}}
\label{t2}
\end{table}
\section{NISQ Implementation of the cryptographic protocol: Results and Challenges} \label{A3}
\begin{figure}[htbp]
    \centering
    
    \begin{subfigure}[b]{0.45\textwidth}
        \centering
        \includegraphics[width=\textwidth]{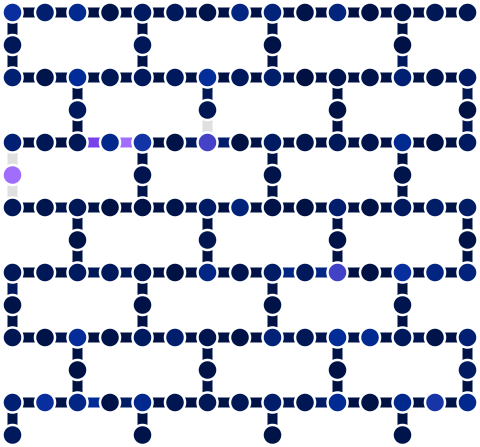}
        \caption{}
        \label{torfig}
    \end{subfigure}
    \hfill 
    \begin{subfigure}[b]{0.49\textwidth}
        \centering
        \raisebox{2.5cm}{ 
            \begin{tabular}{|c|c|}
                \hline
                
                Processor & Heron $r1$ \\ \hline
                Total qubits    & 133 \\ \hline
                Basis Gates    & \shortstack{$id, rz,  sx$, \\$ x, cz$ }\\ \hline
                \shortstack{Circuit layer Operations \\ per second (CLOPs)} & 210000\\ \hline
                Median $T_1$    & 175 $\mu s$   \\ \hline
                Median $T_2$ & 143 $\mu s$\\\hline
                Median readout error & $1.86 \times 10^{-2}$ \\ \hline
                \shortstack{Median single-qubit \\ error} & $0.8 \times 10^{-3}$\\\hline
                \shortstack{Median two-qubit \\ error} & $1.39 \times 10^{-3}$\\\hline
            \end{tabular}
        }
        \caption{}
        \label{tordata}
    \end{subfigure}
    
    \caption{(a) Qubit connectivity map and (b) Technical specifications of \texttt{ibm\_torino} quantum device. The basis (native) gates include the identity ($id$), single-qubit rotations about the $Z$-axis ($rz$), single-qubit rotations of angle $\pi/2$ about the $X$-axis ($sx$), single-qubit rotations of angle $\pi$ about the $X$-axis ($x$), and two-qubit controlled $Z$ ($cz$) gate. }
    \label{ibmtorino}
\end{figure}

Here, we present the results that we obtained via the implementation of the proposed Parrondo's paradox-based cryptographic protocol on a superconducting qubit-based quantum computer, \texttt{ibm\_torino}, a 133-qubit device under the \textit{Heron r1} processor. \textcolor{black}{Fig.~\ref{ibmtorino} (a) shows the qubit connectivity map of \texttt{ibm\_torino}. The device specifications, that is, the median relaxation time ($T_1$), dephasing time ($T_2$), single-qubit, two-qubit errors, readout errors, the native gates (basis gates), and the circuit layer operations per second (CLOPS) which is a, measure of how many layers of a quantum circuit a quantum device can execute per unit of time~\cite{r15, arxiv} are listed in Fig.~\ref{ibmtorino} (b).} The quantum system is accessed via IBM cloud services~\cite{r15}. This section focuses on the realization of communication between Alice and Bob using different quantum state-transfer strategies, namely direct state transfer via only SWAP operations and hybrid schemes combining SWAP and quantum teleportation. We analyze how these strategies are affected by qubit placement, device connectivity, and transpiler optimization levels, and compare their impact on circuit depth and decrypted-message fidelity. The results illustrate the trade-offs between preserving the logical separation of communicating parties and mitigating hardware-induced errors, providing insight into the feasibility of distributed cryptographic protocols on current NISQ devices.

\subsection{Quantum state Transfer with SWAP gates}~\label{C1}
\begin{figure}[h]
    \centering
    \includegraphics[width=1\linewidth]{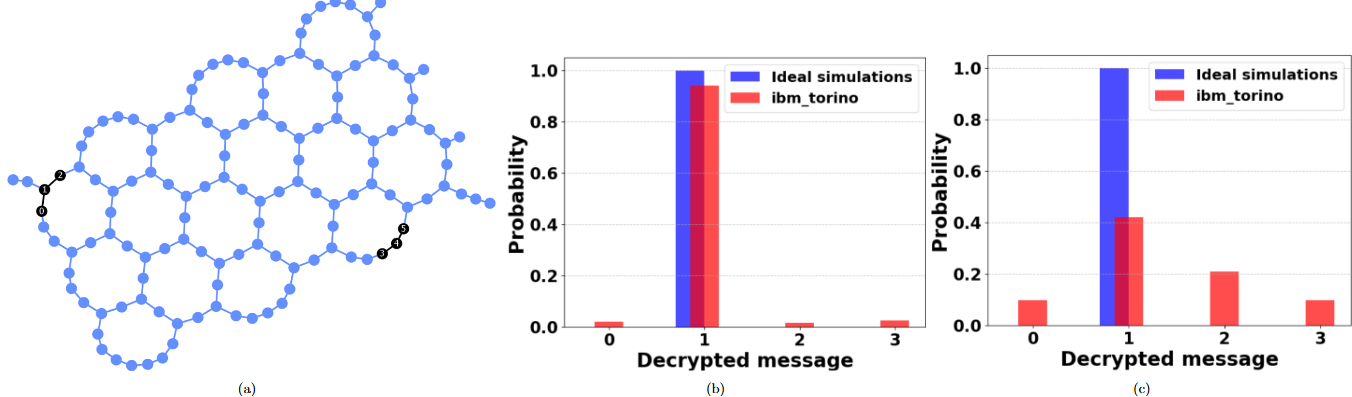}
    \caption{Realization of the quantum circuit shown in Fig.~\ref{totalckt} on the \texttt{ibm\_torino} processor for $10^5$ shots on a 4-cycle graph.\textcolor{black}{(a) Qubit layout of the \texttt{ibm\_torino} processor showing the physical coupling graph. Blue circles represent all available physical qubits on the device; black circles indicate the qubits selected for the protocol, with Alice's qubits ($q_0,q_1,q_2$) mapped to physical qubits (11,12,18) and Bob's qubits ($q_3,q_4,q_5$) mapped to physical qubits (91,95,96). Connecting lines between the circles denote the physical couplings between those qubits.}(b) Probability distribution of the decrypted message $k=1$ obtained by Alice for the initial state $\ket{x}=\ket{0}$ and $\ket{l}=\ket{0}$ at transpiler optimization level 3. (c) Corresponding distribution at optimization level 1.}
    \label{IBM-1}
\end{figure}

In principle, the proposed cryptographic protocol requires Alice and Bob to reside on two spatially separated quantum processors. As such distributed quantum execution is not yet supported on current NISQ platforms, we emulated this setting on a single IBM quantum processor by allocating disjoint and widely separated qubit registers (see, Fig.~\ref{IBM-1} (a)) to represent Alice (position states are encoded in qubits 11, 12 and coin in qubit 18) and Bob (position states are encoded in qubits 91, 95 and coin in qubit 96). The registers were chosen to be as far apart as possible on the device coupling graph to enforce logical separation.

Quantum state transfer between Alice and Bob was implemented through a sequence of SWAP operations across intermediate qubits (see, Algorithm 1 (Fig.~\ref{a1}) in Main). At transpiler optimization level 3 (Fig.~\ref{IBM-1} (b)), aggressive qubit remapping and gate reordering substantially alter the intended circuit structure, effectively reducing the distinguishability between the Alice and Bob modules. Although this optimization reduces nominal circuit depth, it compromises the logical separation required by the protocol and leads to deviations from the expected behavior. The qubits were initially mapped to separate modules (see, Fig.~\ref{IBM-1} (a)), but the transpiler optimization compromises the intended communication structure. In particular, only three qubits corresponding to Alice are ultimately retained by the quantum processor, effectively erasing the logical separation between Alice and Bob and rendering any cryptographic message transfer futile. In contrast, optimization level 1 (Fig.~\ref{IBM-1} (c)) better preserves the logical layout and communication pathways between Alice and Bob, but at the cost of increased circuit depth and a correspondingly decreased fidelity due to higher susceptibility to noise. Fig.~\ref{IBM-1} thus highlights a fundamental trade-off between structural fidelity and noise resilience on current NISQ hardware. A key limitation of this approach is that it compromises the modular structure required by the cryptographic protocol. To address this, we avoid direct qubit state transfer and instead explore communication strategies that preserve the logical modules while enabling information transfer, as discussed in the following subsection.
\subsection{Quantum State Transfer via SWAP and quantum teleportation operations}\label{C2}
\begin{figure}[H]
    \centering
    \includegraphics[width=0.9\linewidth]{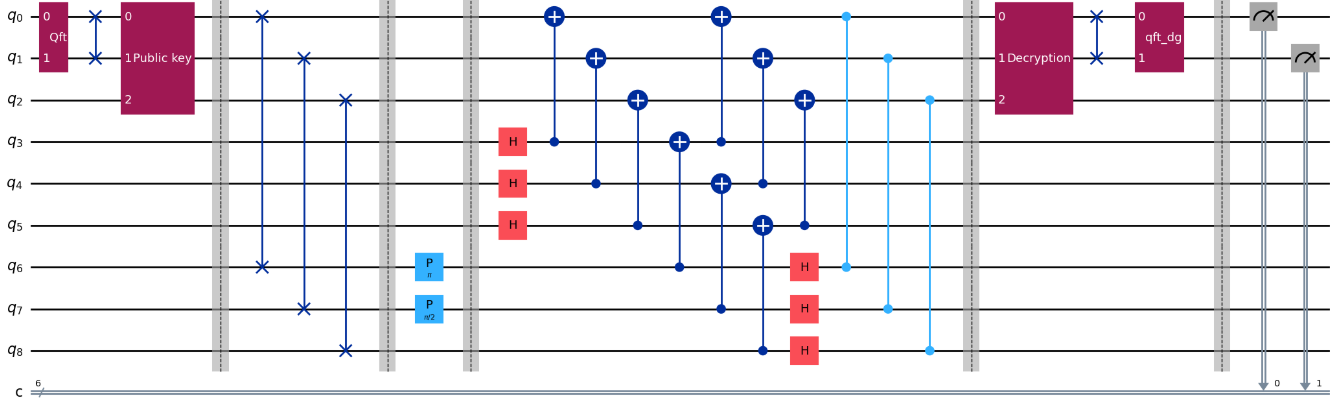}
    \caption
    {Quantum circuit implementing the proposed Parrondo’s paradox–based cryptographic protocol with hybrid state transfer via SWAP followed by quantum teleportation. Qubits $q_0,q_1$ and $q_2$ represent Alice’s position and coin degrees of freedom. The public key generated by Alice with initial state $|000\rangle$ is first transferred to Bob via SWAP operations. Ancilla qubits $q_3,q_4$ and $q_5$ generate entanglement between Alice’s and Bob’s modules for the teleportation stage, while qubits $q_6,q_7$ and $q_8$ constitute Bob’s module. Bob encodes the message $k=1$ on his module, after which the encrypted state is transmitted back to Alice via quantum teleportation, followed by subsequent decryption by Alice.}
    \label{ST-1}
\end{figure}

\begin{figure}[H]
    \centering
    \includegraphics[width=0.9\linewidth]{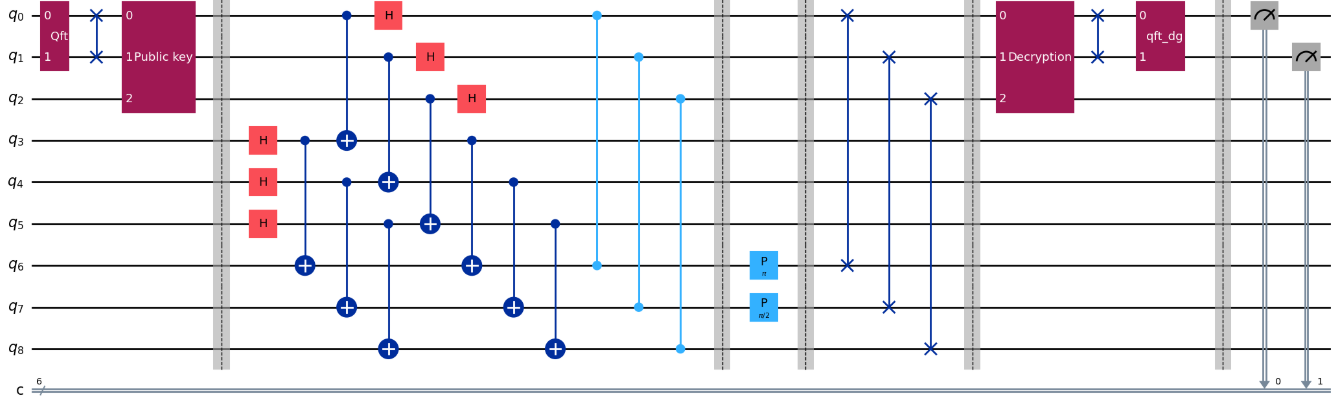}
    \caption{Quantum circuit implementing the proposed Parrondo’s paradox–based cryptographic protocol with hybrid state transfer via quantum teleportation followed by SWAP. Qubits $q_0,q_1$ and $q_2$ represent Alice’s position and coin degrees of freedom. The public key generated by Alice with initial state $|000\rangle$ is first transferred to Bob via quantum teleportation. Ancilla qubits $q_3,q_4$ and $q_5$ generate entanglement between Alice’s and Bob’s modules for the teleportation stage, while qubits $q_6,q_7$ and $q_8$ constitute Bob’s module. Bob encodes the message $k=1$ on his module, after which the encrypted state is transmitted back to Alice via SWAP operations followed by subsequent decryption by Alice.}
    \label{ST-2}
\end{figure}

To partially mitigate the issue (no effective distinction between the communicating parties) incurred in the previous section, we explored a modified communication strategy in which one stage of state transfer is employed via SWAP operations and the other with quantum teleportation between Alice and Bob. Conceptually, this hybrid approach is closer to a distributed protocol, as teleportation avoids long chains of SWAP operations and reduces qubit state transfer across the device. To implement the hybrid communication protocol, we modify Algorithm 1 (Fig.~\ref{a1}) and present two variants: (i) quantum state transfer via SWAP operations followed by quantum teleportation, and (ii) quantum teleportation followed by a SWAP-based transfer. For both cases, we provide the corresponding Algorithms 4 (Fig.~\ref{a3}) and 5 (Fig.~\ref{a4}) and explicit Qiskit circuit realizations (Figs.~\ref{ST-1} and \ref{ST-2}) to analyze the impact of operation ordering on preservation of logical modularity. We extend the circuit by three ancilla qubits that are used to generate entangled Bell pairs, allowing the quantum states to be teleported, thereby mimicking inter-module quantum communication. We now proceed to discuss the implementation of this hybrid protocol on \texttt{ibm\_torino} using different qubit layouts and connectivity between the modules.

\begin{figure}[H]
\caption{\textbf{Algorithm 4}: Parrondo's Paradox DTQW-based Cryptographic Protocol with state transfer first via SWAP then Teleport (see, Fig.~\ref{ST-1})}
\label{a3}
\centering
\fbox{%
\begin{minipage}{0.95\columnwidth}
\small
\begin{algorithmic}[1]
\State \textbf{Input:} Pre-shared private key inputs: Coin operators $A$, $B$ generating chaotic DTQW dynamics, time steps $t$, initial coin $\ket{l}$ and position state $\ket{x}$ of walker
\State Design the circuit with 9 qubits: Alice 3 qubits $q_0q_1q_2$, Bob 6 qubits: 3 qubits $q_3q_4q_5$ to generate entanglement, 3 system qubits $q_6q_7q_8$; Position qubits: $q_0q_1$, $q_6q_7$, Coin qubits: $q_2$ and $q_8$ 
\Statex \textbf{Public Key Generation by Alice}
\State Initialize $q_0q_1q_2$
\State Apply QFT on position qubits of Alice $q_1q_0$ 
\State Apply SWAP on $q_1q_0$ to get the correct output 
\For{$i = 0$ to $2$}
    \State Perform the coin operation B on $q_2$
    \State Perform the phase rotation $P(-\pi)$ on $q_0$  
    \State Perform the phase rotation $P(-\pi / 2)$ on $q_1$
    \If{$\ket{q_2}$ is in $\ket{1}$}
        \State Perform the phase rotation $P(\pi)$ on $q_1$
    \EndIf
\EndFor
\Statex\textbf{Public Key transfer from Alice to Bob via SWAP}
\State Apply the SWAP gates to swap Alice's qubits with Bob's 
\Statex \textbf{ Message Encryption by Bob}
\State Encrypt the message via $T_k^D$ operator on Bob's position qubits $q_6q_7$
\Statex \textbf{Message transfer from Bob to Alice via quantum teleportation} 
\State Generate Bell pairs $\frac{\ket{00}+\ket{11}}{\sqrt{2}}$ with qubits $q_3q_0$, $q_4q_1$, $q_5q_2$  
\State Apply CNOT on qubits $q_3,q_4,q_5$ with controls $q_6,q_7,q_8$ respectively.
\State Apply Hadamard gates on qubits $q_6,q_7,q_8$.
\State Apply CNOT on qubits $q_0,q_1,q_2$ with $q_3,q_4,q_5$ as controls.
\State Apply CZ gates on qubits $q_0,q_1,q_2$ with $q_6,q_7,q_8$ as controls.
\Statex \textbf{Message decryption by Alice}
\For{$i = 0$ to $18$} 
    \If{$i \bmod 4 = 0$ or $i \bmod 4 = 1$} 
        \State Perform the coin operation $A$ on $q_2$
    \Else
        \State Perform the coin operation $B$ on $q_2$
    \EndIf
    \State Perform the phase rotation $P(-\pi)$ on $q_0$  
    \State Perform the phase rotation $P(-\pi / 2)$ on $q_1$
    \If{$\ket{q_2}$ is in $\ket{1}$}
        \State Perform the phase rotation $P(\pi)$ on $q_1$
    \EndIf
\EndFor
\State Apply SWAP on $q_1q_0$ to get the correct output 
\State Apply IQFT on position qubits $q_1q_0$
\State Alice measures her position qubits $q_1q_0$
    \end{algorithmic}
    \end{minipage}%
}
\end{figure}

\begin{figure}[H]
\caption{\textbf{Algorithm 5}: Parrondo's Paradox DTQW-based Cryptographic Protocol with state transfer first via teleport then SWAP(see, Fig.~\ref{ST-2})}
\label{a4}
\centering
\fbox{%
\begin{minipage}{0.95\columnwidth}
\small
\begin{algorithmic}[1]
\State \textbf{Input:} Pre-shared private key inputs: Coin operators $A$, $B$ generating chaotic DTQW dynamics, time steps $t$, initial coin $\ket{l}$ and position state $\ket{x}$ of walker
\State Design the circuit with 9 qubits: Alice 6 qubits: 3 system qubits $q_0q_1q_2$, 3 qubits $q_3q_4q_5$ to generate entanglement, Bob 3 qubits $q_6q_7q_8$; Position qubits: $q_0q_1$, $q_6q_7$, Coin qubits: $q_2$ and $q_8$ 
\Statex \textbf{Public Key Generation by Alice}
\State Initialize $q_0q_1q_2$
\State Apply QFT on position qubits of Alice $q_1q_0$
\State Apply SWAP on $q_1q_0$ to get the correct output 
\For{$i = 0$ to $2$}
    \State Perform the coin operation B on $q_2$
    \State Perform the phase rotation $P(-\pi)$ on $q_0$  
    \State Perform the phase rotation $P(-\pi / 2)$ on $q_1$
    \If{$\ket{q_2}$ is in $\ket{1}$}
        \State Perform the phase rotation $P(\pi)$ on $q_1$
    \EndIf
\EndFor
\Statex\textbf{Public Key transfer from Alice to Bob via quantum teleportation}
\State Generate Bell pairs $\frac{\ket{00}+\ket{11}}{\sqrt{2}}$ with qubits $q_3q_6$, $q_4q_7$, $q_5q_8$  
\State Apply CNOT on qubits $q_3,q_4,q_5$ with controls $q_0,q_1,q_2$ respectively.
\State Apply Hadamard gates on qubits $q_0,q_1,q_2$.
\State Apply CNOT on qubits $q_6,q_7,q_8$ with $q_3,q_4,q_5$ as controls.
\State Apply CZ gates on qubits $q_6,q_7,q_8$ with $q_0,q_1,q_2$ as controls.
\Statex \textbf{ Message Encryption by Bob}
 \State Encrypt the message via  $T_k^D$ operator on Bob's position qubits $q_6q_7$
\Statex \textbf{Message transfer from Bob to Alice via SWAP} 
\State Apply the SWAP gates to swap Bob's qubits with Alice's
\Statex \textbf{Message decryption by Alice}
\For{$i = 0$ to $18$} 
    \If{$i \bmod 4 = 0$ or $i \bmod 4 = 1$} 
        \State Perform the coin operation $A$ on $q_2$
    \Else
        \State Perform the coin operation $B$ on $q_2$
    \EndIf
    \State Perform the phase rotation $P(-\pi)$ on $q_0$  
    \State Perform the phase rotation $P(-\pi / 2)$ on $q_1$
    \If{$\ket{q_2}$ is in $\ket{1}$}
        \State Perform the phase rotation $P(\pi)$ on $q_1$
    \EndIf
\EndFor
\State Apply SWAP on $q_1q_0$ to get the correct output 
\State Apply IQFT on position qubits $q_1q_0$
\State Alice measures her position qubits $q_1q_0$
    \end{algorithmic}
    \end{minipage}%
}
\end{figure}
\subsubsection{\textbf{Layout 1: Alice and Bob widely separated}}~\label{C21}
\begin{figure}[H]
    \centering
    \includegraphics[width=1\linewidth]{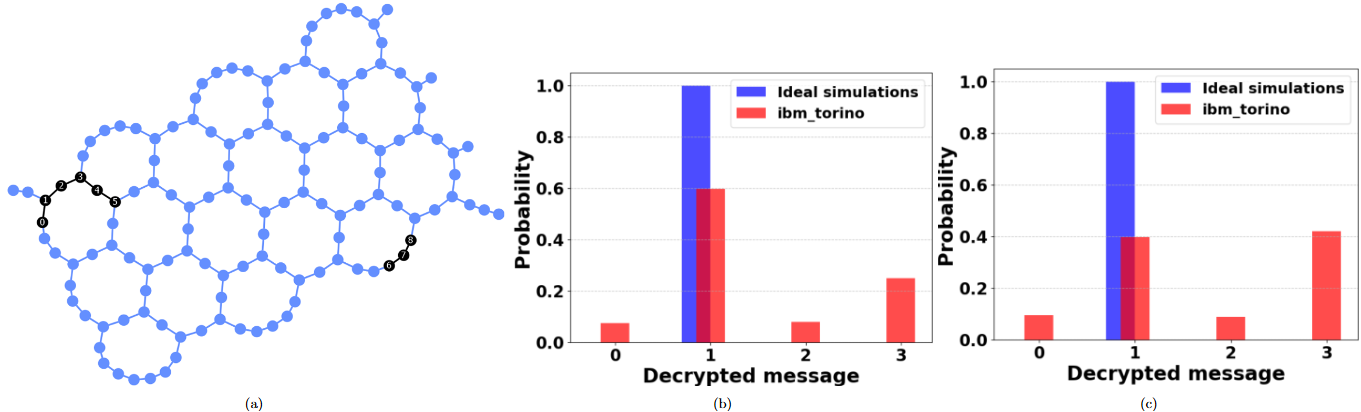}
    \caption{Hybrid state-transfer implementation of the Qiskit circuits on the \texttt{ibm\_torino} processor for $10^5$ shots at transpiler optimization level 3. \textcolor{black}{(a) Qubit layout on the \texttt{ibm\_torino} processor showing the physical coupling graph. Blue circles represent all available physical qubits on the device; black circles indicate the qubits selected for the protocol, with Alice's qubits ($q_0,q_1,q_2$) mapped to physical qubits (11,12,18), entanglement-generation ancilla qubits on ($q_3,q_4,q_5$) mapped to (31,30,29), and Bob's qubits ($q_6,q_7,q_8$) mapped to physical qubits (91,95,96). Connecting lines between the circles denote the physical couplings between those qubits.}
    (b) Probability distribution of the decrypted message $k=1$ obtained using the state transfer scheme with first teleportation, then SWAP (Fig. \ref{ST-2} as the Qiskit circuit), and (c) first SWAP then teleport (Fig. \ref{ST-1} as the Qiskit circuit).}
    \label{TS&ST}
\end{figure}

For this case, we retain the same modular layout shown in Fig.~\ref{IBM-1} (a) and introduce additional qubits to generate Bell pairs required for the teleportation stage of the hybrid protocol. The qubit layout is depicted in Fig~\ref{TS&ST} (a). The circuits shown in Figs.~\ref{ST-1} and \ref{ST-2} are realized in \texttt{ibm\_torino} at optimization level 3 with qubits $q_0q_1q_2$ encoded in physical qubits $11,12,18$, $q_3q_4q_5$ in $31,30,29$ and $q_6q_7q_8$ in $91,95,96$. 

Figs.~\ref{TS&ST} (b) and (c) show the probability distribution of the decrypted message for $k=1$ obtained using the hybrid state-transfer protocol. When the circuits are realized in \texttt{qiskit\_aer} under ideal (noise-free) conditions, the decrypted message is recovered successfully with unit probability for both schemes, confirming the correctness of the circuit construction. When implemented on the NISQ hardware, although the correct decrypted message is still recovered as the most probable outcome, the fidelity is reduced ($\approx$ 59.6\% for (b) and $\approx$ 39\% for (c)), indicating partial degradation of the protocol due to the increase in system size together with hardware noise and gate imperfections. Importantly, the hybrid communication strategy preserves the intended communication structure, namely the logical distinction between Alice and Bob, which is otherwise compromised in purely SWAP-based implementations. A comparison of panels (b) and (c) further reveals that the ordering of the communication primitives affects the observed fidelity, with differences arising from how errors accumulate during SWAP operations and teleportation. Motivated by these observations, we examine alternative qubit layouts in the following section to assess whether performing SWAP followed by teleportation is more favorable than teleportation followed by SWAP on current-generation quantum hardware, where qubits and the two-qubit gates remain imperfect.

\subsubsection{\textbf{Layout 2: Alice and Bob moderately separated}}~\label{C22}
\begin{figure}[H]
    \centering
    \includegraphics[width=1\linewidth]{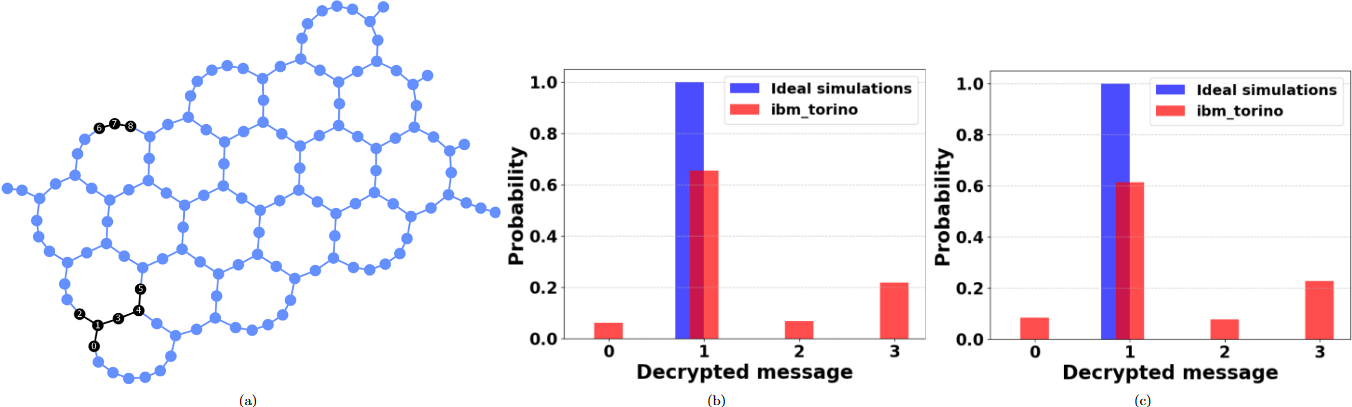}
    \caption{Hybrid state-transfer implementation of the Qiskit circuits on the \texttt{ibm\_torino} processor for $10^5$ shots at transpiler optimization level 3. \textcolor{black}{(a) Qubit layout of the \texttt{ibm\_torino} processor showing the physical coupling graph. Blue circles represent all available physical qubits on the device; black circles indicate the qubits selected for the protocol, with Alice's qubits ($q_0,q_1,q_2$) mapped to physical qubits (3,4,5), entanglement-generation ancilla qubits on ($q_3,q_4,q_5$) mapped to (16,23,24), and Bob's qubits ($q_6,q_7,q_8$) mapped to physical qubits (37,52,51). Connecting lines between the circles denote the physical couplings between those qubits.}
    (b) Probability distribution of the decrypted message $k=1$ obtained using the state transfer scheme with first teleportation, then SWAP (Fig. \ref{ST-2} as the Qiskit circuit), and (c) first SWAP then teleport (Fig. \ref{ST-1} as the Qiskit circuit).}
    \label{ST-1_result}
\end{figure}

\begin{figure}[H]
    \centering
    \includegraphics[width=1\linewidth]{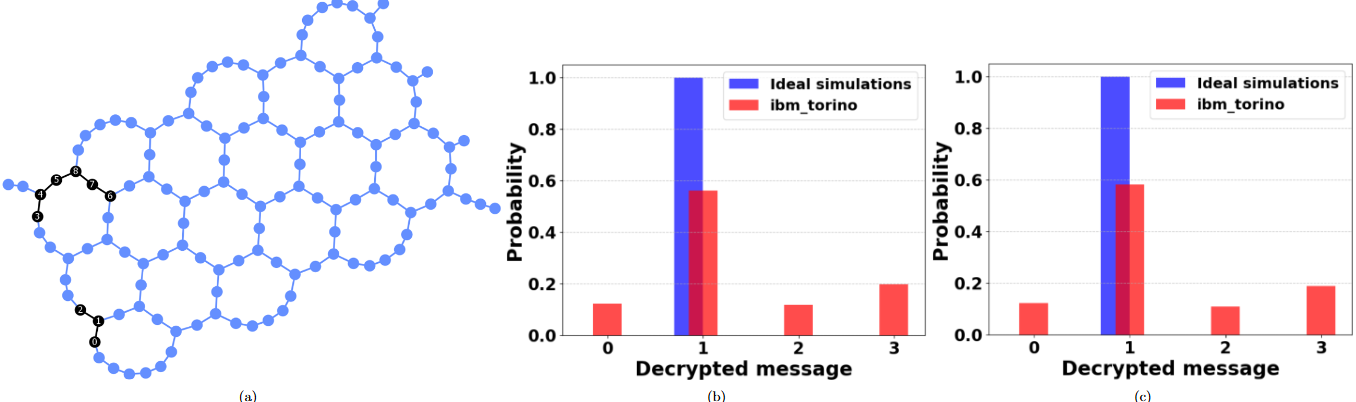}
    \caption{Hybrid state-transfer implementation of the Qiskit circuits on the \texttt{ibm\_torino} processor for $10^5$ shots at transpiler optimization level 3. \textcolor{black}{(a) Qubit layout of the \texttt{ibm\_torino} processor showing the physical coupling graph. Blue circles represent all available physical qubits on the device; black circles indicate the qubits selected for the protocol, with Alice's qubits ($q_0,q_1,q_2$) mapped to physical qubits (3,4,5), entanglement-generation ancilla qubits on ($q_3,q_4,q_5$) mapped to (11,12,18), and Bob's qubits ($q_6,q_7,q_8$) mapped to physical qubits (31,30,29). Connecting lines between the circles denote the physical couplings between those qubits.}
    (b) Probability distribution of the decrypted message $k=1$ obtained using the state transfer scheme with first teleportation, then SWAP (Fig. \ref{ST-2} as the Qiskit circuit), and (c) first SWAP then teleport (Fig. \ref{ST-1} as the Qiskit circuit).}
    \label{ST-2_result}
\end{figure}

In this section, we consider two distinct reduced-separation configurations, shown in Figs.~\ref{ST-1_result} (a) and \ref{ST-2_result} (a). Although both layouts reduce the logical distance between Alice and Bob, they differ in the specific physical qubits and connectivity patterns on the hardware, leading to contrasting performance trends.
Fig.~\ref{ST-1_result} (a) shows the qubit layout, where Alice and Bob are placed closer on the \texttt{ibm\_torino} coupling graph while still maintaining distinct logical modules with qubits $q_0q_1q_2$ encoded in physical qubits $3,4,5$, $q_3q_4q_5$ in $16,23,24$, and $q_6q_7q_8$ in $37,52,51$. 

Figs.~\ref{ST-1_result} (b) and (c) present the probability distributions of the decrypted message for $k=1$ obtained using the hybrid state-transfer protocol with the layout depicted in Fig.~\ref{ST-1_result} (a). While ideal simulations yield unit-probability recovery of the correct message, execution on IBM hardware results in a broadened distribution. A direct comparison of panels (b) and (c) shows that the SWAP-then-teleport strategy yields lower fidelity ($\approx$ 61.3\%) than the teleport-then-SWAP strategy($\approx$ 65.3\%), underscoring how errors associated with Bell-pair generation, measurements, and additional two-qubit gates accumulate in the circuit. To examine whether the SWAP-then-teleport ordering can yield improved performance under a different qubit configuration, we next consider an alternative layout, given in Fig.~\ref{ST-2_result} that modifies the choice of physical qubits and their connectivity on the device coupling graph. 

Fig.~\ref{ST-2_result} (a) shows the qubit layout, where Alice and Bob are placed closer on the \texttt{ibm\_torino} coupling graph while still maintaining distinct logical modules with qubits $q_0q_1q_2$ encoded in physical qubits $3,4,5$, $q_3q_4q_5$ in $11,12,18$, and $q_6q_7q_8$ in $31,30,29$. 

Figs.~\ref{ST-2_result} (b) and (c) show the probability distributions of the decrypted message for $k=1$ obtained using the hybrid state-transfer scheme implemented on \texttt{ibm\_torino} with the layout depicted in Fig.~\ref{ST-2_result} (a). A direct comparison of panels (b) and (c) reveals that the SWAP-then-teleport (fidelity $\approx$ 58\%) strategy yields higher fidelity than the teleport-then-SWAP (fidelity $\approx$ 56\%) strategy, for this specific choice of physical qubits, indicating that the relative ordering of communication primitives remains strongly conditioned on the underlying qubit layout and connectivity. However, the overall fidelity in this layout (see, Fig.~\ref{ST-2_result} (a)) is reduced in comparison to the previous reduced-separation configuration (see, Fig.~\ref{ST-1_result} (a)). These results further highlight the strong dependence of protocol performance on choice of physical qubits, and demonstrate that the optimal ordering of SWAP and teleportation operations is inherently hardware dependent.\\Motivated by the strong dependence of performance on qubit choice and connectivity, we next consider a qubit layout that exploits a fully connected architecture, allowing us to minimize routing overhead and further isolate the impact of hardware noise from connectivity-induced errors.

\subsubsection{\textbf{Layout 3: Fully connected modules}}~\label{C23}
\begin{figure}[h]
    \centering
    \includegraphics[width=1\linewidth]{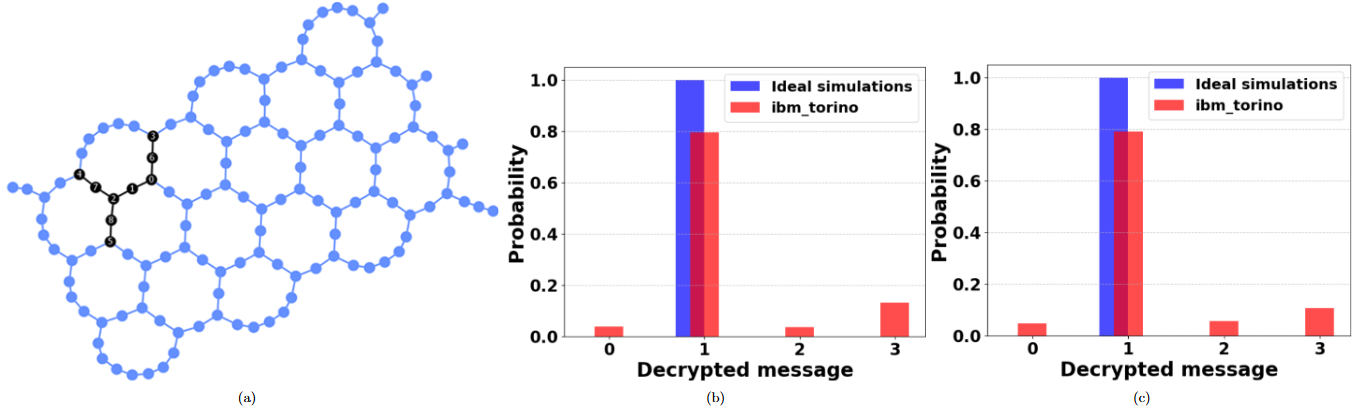}
    \caption{Hybrid state-transfer implementation of the Qiskit circuits of the \texttt{ibm\_torino} processor for $10^5$ shots at transpiler optimization level 3. \textcolor{black}{(a) Qubit layout on the \texttt{ibm\_torino} processor showing the physical coupling graph. Blue circles represent all available physical qubits on the device; black circles indicate the qubits selected for the protocol, with Alice's qubits ($q_0,q_1,q_2$) mapped to physical qubits (28,29,30), entanglement-generation ancilla qubits on ($q_3,q_4,q_5$) mapped to (31,36,47), and Bob's qubits ($q_6,q_7,q_8$) mapped to physical qubits (48,49,50). Connecting lines between the circles denote the physical couplings between those qubits.}
    (b) Probability distribution of the decrypted message $k=1$ obtained using the state transfer scheme with first teleportation, then SWAP (Fig. \ref{ST-2} as the Qiskit circuit), and (c) first SWAP then teleport (Fig. \ref{ST-1} as the Qiskit circuit).}
    \label{ST-3_result}
\end{figure}

Figs.~\ref{ST-3_result} (a) shows the qubit layout, where Alice and Bob are fully connected on the \texttt{ibm\_torino} coupling graph with qubits $q_0q_1q_2$ encoded in physical qubits $28,29,30$, $q_3q_4q_5$ in $31,36,47$, and $q_6q_7q_8$ in $48,49,50$. 
Fig.~\ref{ST-3_result} (b) and (c) show the probability distribution of the decrypted message for $k=1$ obtained using the hybrid state-transfer protocol when Alice and Bob are embedded within a fully connected register architecture on the \texttt{ibm\_torino} processor. In the ideal simulations, the decrypted message is recovered with unit probability for both hybrid variants. On hardware, the fully connected layout leads to a marked improvement in fidelity compared to the previous layouts, as reduced routing overhead and improved connectivity suppress error accumulation from long SWAP chains. Although hardware noise and gate imperfections still broaden the output distribution, the correct decrypted message exhibits a significantly higher fidelity ($\approx$ 79\%) in both panels. Overall, these results demonstrate that access to highly connected qubit modules plays a crucial role in preserving fidelity on current-generation quantum hardware.

Taken together, these results demonstrate that the principal obstacle is not the cryptographic protocol itself, but the lack of true hardware-level modularity and inter-processor connectivity in current NISQ platforms. When implemented on a single quantum processor, the protocol either collapses into an effectively single-module computation due to transpiler optimizations or suffers severe fidelity loss due to excessive circuit depth and noise when modularity is enforced manually. The results of the NISQ implementation of the proposed cryptographic protocol, comparing different state-transfer strategies and qubit layouts used for the Alice and Bob modules, are summarized in Table~\ref{t3}. It highlights how the choice of communication method, circuit depth (number of sequential operations of gate layers), and module connectivity jointly influence decryption fidelity.

Together, these observations underscore the present limitations of faithfully realizing spatially separated cryptographic protocols on a single quantum processor. They motivate our simulation-based, proof-of-principle validation, while illustrating the practical challenges that must be addressed before such protocols can be reliably deployed on current NISQ hardware.

\begin{table}[H]
\centering
    \renewcommand{\arraystretch}{1.1}
\begin{tabular}{|l|c|c|c|c|c|c|c|c|}
\hline
\textbf{\shortstack{State transfer\\ Protocol (Opt.level)}}&\textbf{\shortstack{Qubits \\required}} &\textbf{\shortstack{Transpiled\\ qubits}} &\textbf{\shortstack{Distinction \\between \\Modules}}&\textbf{Module} & \textbf{\shortstack{Physical\\ Qubit \\Layout}} &\textbf{Depth }&\textbf{\shortstack{Hellinger \\Fidelity}}&\textbf{Remarks}\\
\hline
{only SWAPs (3)} &6 &3 & Lost & \shortstack{Widely \\Separated}  &\shortstack{11,12,18,\\91,95,96} & 36 & 94\% & \shortstack{Implementation \\invalid}\\
\hline
{only SWAPs (1)} &6 &45 & Preserved & \shortstack{Widely \\Separated}  &\shortstack{11,12,18,\\91,95,96} & 350 & 41\% & \shortstack{Fidelity below \\threshold}\\
\hline
{Teleport then SWAP (3)} &9 &45 & Preserved&\shortstack{Widely \\Separated}   &\shortstack{11,12,18,\\31,30,29,\\91,95,96} & 188 & $\approx$ 59\% & Acceptable\\
\hline
{SWAP then Teleport (3)} &9 &45 & Preserved & \shortstack{Widely \\Separated}   &\shortstack{11,12,18,\\31,30,29,\\91,95,96} & 204 & $\approx$ 39\% & \shortstack{Fidelity below\\ threshold}\\
\hline
{Teleport then SWAP (3)} &9 &24 & Preserved &\shortstack{Moderately \\Separated}  &\shortstack{26,27,28,\\29,36,48,\\60,59,72} & 193 & $\approx$ 29.4\% & \shortstack{Fidelity below\\ threshold} \\
\hline
{SWAP then Teleport (3)} &9 &24 & Preserved & \shortstack{Moderately \\Separated}  &\shortstack{26,27,28,\\29,36,48,\\60,59,72} & 198 & $\approx$ 32.9\% &\shortstack{Fidelity below\\ threshold}\\
\hline
{SWAP then Teleport (3)} &9  &17 & Preserved & \shortstack{Moderately \\Separated} &\shortstack{3,4,5,\\11,12,18,\\31,30,29} & 139 & $\approx$ 58\% & Acceptable\\
\hline
{Teleport then SWAP (3)} &9 &17 & Preserved & \shortstack{Moderately \\Separated} &\shortstack{3,4,5,\\11,12,18,\\31,30,29} & 160 & $\approx$ 56\% &Acceptable\\
\hline
{Teleport then SWAP (3)} &9 &17 & Preserved & \shortstack{Moderately \\Separated}&\shortstack{31,30,29,\\11,12,18,\\3,4,5} & 137 & $\approx$ 67\% &\shortstack{Improved \\performance}\\
\hline
{SWAP then teleport (3)} &9 &26 & Preserved& \shortstack{Moderately \\Separated}&\shortstack{3,4,5,\\16,23,24,\\37,52,51} & 161 & $\approx$ 61.3\%&\shortstack{Improved \\performance} \\
\hline
{Teleport then SWAP (3)} &9 &26 & Preserved & \shortstack{Moderately \\Separated} &\shortstack{3,4,5,\\16,23,24,\\37,52,51} & 169 & $\approx$ 65.342\% &\shortstack{Improved \\performance}\\
\hline
{SWAP then Teleport (3)} &9 &26 & Preserved & \shortstack{Moderately \\Separated} &\shortstack{37,52,51,\\3,4,5,\\16,23,24} & 184 & $\approx$ 61.9\% &\shortstack{Improved \\performance}\\
\hline
{Teleport then SWAP (3)} &9 &9 & Preserved &\shortstack{Fully \\Connected} &\shortstack{28,29,30,\\31,36,47,\\48,49,50} & 100 & $\approx$ 79\% &Optimal\\
\hline
{SWAP then Teleport (3)} &9 &9 & Preserved & \shortstack{Fully \\Connected} &\shortstack{28,29,30,\\31,36,47,\\48,49,50} & 98 & $\approx$ 79\% &Optimal\\
\hline
\end{tabular}
\caption{Comparison of the proposed cryptographic protocol implemented on \texttt{ibm\_torino} using different state-transfer methods. Note: The first SWAP-only protocol (optimization level 3 ) is not a valid implementation, as the distinction between Alice and Bob is lost. Rest all approaches remain valid, with differing fidelities.}
\label{t3}
\end{table}
\section{\textcolor{black}{Security of the proposed Parrondo-based quantum cryptographic protocol}}~\label{A5}
\textcolor{black}{The security of the proposed Parrondo-based quantum cryptographic protocol relies on the difficulty of reconstructing the quantum walk parameters that generate the public key ($PuK$) and the corresponding private key ($PvK$) required for decryption~\cite{DKP, QPKE}. An adversary (Eve) attempting to compromise the protocol must determine the parameters of the set $\{W,t,l,x\}$ that define the public key (see Eq.(\ref{e33})), namely the chaotic walk operator ($W$), the evolution time ($t$), and the initial walker state $(l,x)$, from an intercepted public-key state. Since these parameters, along with the decryption operator ($G$), collectively determine the private key ($PvK$), the security of the protocol can be quantified through the probability that an eavesdropper successfully infers the underlying DTQW parameters. The chaotic walk operator is selected from the set, $\mathcal{W} = \{ W_i = W(s_i, \gamma_i, \delta_i) | i \in [1, D]\}$, where $D$ can be taken to be arbitrarily large, since the walk operator depends continuously on the coin parameters $s_i, \gamma_i, \delta_i$. Furthermore, the evolution time is selected from the set, $t \in \{t_1, t_2,....t_{max}\}$, while the initial walker state is specified by the coin state $l\in\{0,1\}$ and the position state $x\in\{0,1,\ldots,K-1\}$ on a $K$-cycle graph. Consequently, yielding $2KD|T|$ possible configurations, with $|T|$ being the cardinality of the time steps. Since Eve does not know the parameters used to generate the public key, she cannot associate the intercepted state with a unique DTQW evolution. Instead, from her perspective, the parameters of the public key (coin and position) are drawn from an ensemble of possible states, which belong to a Hilbert space of dimension $d = 2K$. Eve may perform a measurement on the intercepted state and obtain information about a particular degree of freedom, such as the coin or position state. However, a single measurement outcome does not uniquely determine the chaotic walk operator or the time steps. Hence, Eve's probability of correctly reconstructing the complete public-key parameter set remains negligible. Eve's best strategy is effectively reduced to guessing the parameters from the set of all admissible configurations. The probability that Eve correctly guesses the public-key parameters scales as \begin{equation}
    \mathcal{P}^{PuK}_{E} = \frac{1}{2KD|T|}.
\end{equation} Since $D$ is arbitrarily large due to the continuous parameter space of the walk operator and $t_{max}$ can be increased, the probability of successfully reconstructing the public-key parameters, i.e.,  $\mathcal{P}^{PuK}_E \rightarrow 0$.} \\
\textcolor{black}{The private key contains, in addition to the public-key parameters, the decryption operator $G$ required to satisfy the relation $GW^t = I$ and recover the encrypted message $k$. Let $G$ be chosen from the set $\mathcal{G} = \{G_i | i \in [1, N_G]\}$, then the possible number of private-key configurations is $2KN_GD|T|$. Accordingly, the probability that Eve guesses the private key scales as \begin{equation}
    \mathcal{P}^{PvK}_{E} = \frac{1}{2KN_GD|T|}.
\end{equation} Since $N_G$ grows with the number of possible chaotic 
operators and their combinations in the Parrondo sequence, the probability of successfully guessing the private key is lower than that for the public key. Therefore, the probability of successfully reconstructing either the public key or the private key from an intercepted public-key state is negligible. As an illustrative example, let us consider $D = 10^6, |T| = 100$ with $K =4$, the number of possible configurations becomes $8 \times 10^8$, yielding $\mathcal{P}_E^{PK} \approx 10^{-9}$. If the private decryption operator is chosen from $N_G = 10^6$ admissible sequences, the private-key space increases to $8 \times 10^{14}$, giving $\mathcal{P}_E^{PvK} \approx 10^{-15}$. These probabilities are negligible and decrease further with increasing $D, |T|$, or $N_G$. Consequently, an adversary attempting to reconstruct either the public key or the private key from an intercepted public-key state faces an exponentially large search space, making successful key recovery computationally and physically infeasible. Further, Eve is detected if she attempts to intercept the communication only once, causing the legitimate parties to abort the protocol and restart the communication process (see, Sec.~\ref{S5}).}
\section{\textcolor{black}{Implementation of the cryptographic protocol with Parrondo sequence $A'A'B'B'B'B'...$ on a 4-cycle graph}}~\label{A4}
\begin{figure}[H]
    \centering
    \includegraphics[width=1\linewidth]{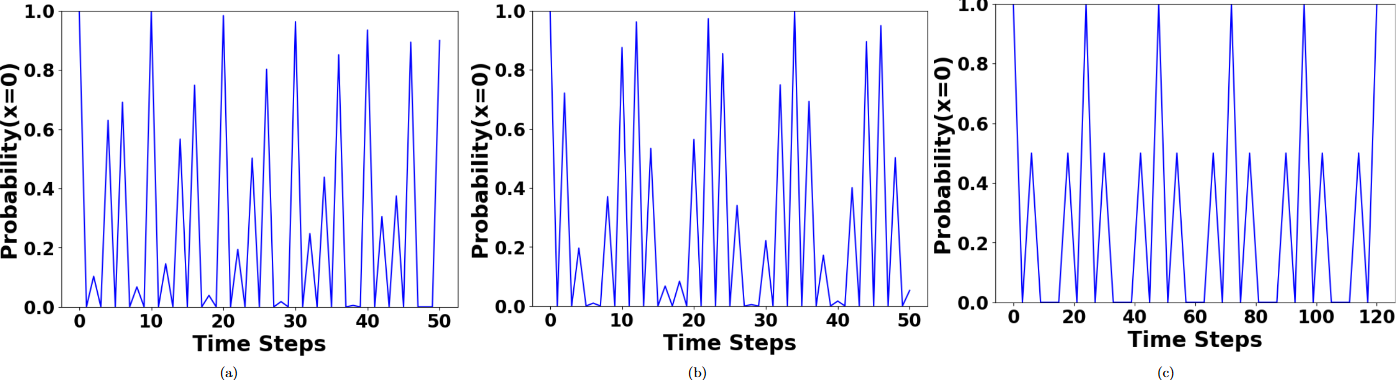}
    \caption{\textcolor{black}{Probability at the initial site x = 0 vs time steps for Parrondo sequences (a) $A'A'A'A'A'A'...$, (b) $B'B'B'B'B'B'...$ up to 50 time steps yielding chaotic quantum walks, and (c) $A'A'B'B'B'B'...$ up to 120 time steps yielding periodic dynamics on a 4-cycle graph.}}
    \label{f26}
\end{figure}

\textcolor{black}{Herein, we demonstrate that the cryptographic protocol discussed in the main text is not restricted to a single Parrondo sequence. In particular, we illustrate that the deterministic sequence $A'A'B'B'B'B'...$, constructed from the chaotic coin operators, generates periodic dynamics on a 4-cycle graph. This provides an additional example of order emerging from the combination of individually chaotic quantum walks and further illustrates the robustness of the Parrondo mechanism underlying the protocol. \\ The analysis presented here closely follows the approach developed in Ref.~\cite{panda}, adapted to the present sequence. We consider three distinct unitary operators (see, Eq.~(\ref{e5})), $A' = W(s_1, \gamma_1, \delta_1)$, $B' = W(s_2, \gamma_2, \delta_2)$, and $P = W(s, \gamma, \delta)$. The coin-evolution operators, $A'$ and $B'$, following sequences $A'A'A'...$ and $B'B'B'...$ yield chaotic quantum walks, while the operator $P$ in sequence $PPP...$ yields an ordered walk with period T. To investigate whether the combination of chaotic operators $A'$ and $B'$ in the sequence $A'A'B'B'B'B'...$ generates periodic dynamics, we start by calculating the eigenvalues of the matrix $MA'A'B'B'B'B'M^\dagger$, where $M$ and $M^\dagger$ are QFT and IQFT matrices defined in Eq.~(\ref{q}) for the 4-cycle graph. The matrix $A'A'B'B'B'B'$ is circulant and hence follows the form defined in Eq.~(\ref{e5}), which is further block diagonalized by the Fourier matrices, given as, \begin{equation}
    W_{K} = \begin{pmatrix}
        W_{K,0} & 0 &... &0\\0 & W_{K,1}&...&0\\
        \vdots &\vdots &\ddots & \vdots\\ 0& 0& ... &W_{K,K-1},
        \end{pmatrix} 
\end{equation} where each block $W_{K,l}$ is a $2\times2$ matrix as defined in Eq.(~\ref{e5}). For a 4-cycle graph, there will be four diagonal blocks~\cite{panda} $W_{4,0},W_{4,1},W_{4,2},W_{4,3}$ and the sum of eigenvalues corresponding to diagonal block $W_{4,1}$ is, 
\begin{equation}
    \alpha^{A'A'B'B'B'B'}_{4,1} = \frac{\alpha^{A'A'B'B'B'B'+}_{4,1} + \alpha^{A'A'B'B'B'B'-}_{4,1}}{2} = 1-8s_2 + 8(2s_2-1)\sqrt{s_1s_2(1-s_1)(1-s_2)} + 8s_2^2 - 2s_1+16s_1s_2(1-s_2).
\end{equation}}
\textcolor{black}{For simplicity, we have considered the phases, $\gamma_1=\gamma_2 = \gamma = \delta_1 = \delta_2 = \delta = 0$. Since our objective is to check whether the sequence $A'A'B'B'B'B'..$ leads to periodicity, we evaluate the eigenvalues for the matrix $MPPPPPPM^\dagger$, where $P$ generates a periodic walk. The sum of the eigenvalues for the diagonal block $W_{4,1}$ is, 
\begin{equation}
    \alpha^{PPPPPP}_{4,1} = \frac{\alpha^{PPPPPP+}_{4,1} + \alpha^{PPPPPP-}_{4,1}}{2} = 1- 2s(3 - 4s)^2.
\end{equation}}
\textcolor{black}{If $A'A'B'B'B'B'..$ were to generate periodic dynamics, its eigenvalues must match with those obtained from $PPPPPP$, i.e., \begin{equation}
    \alpha^{PPPPPP}_{4,1} = \alpha^{A'A'B'B'B'B'}_{4,1}.
\end{equation}}
\textcolor{black}{\begin{subequations}\label{condition}
\begin{align}
4s_2 - 4(2s_2-1)\sqrt{s_1s_2(1-s_1)(1-s_2)} - 4s_2^2 
+ s_1 - 8s_1s_2(1-s_2) &= s(3-4s)^2,
\label{conditiona} \\
(1-8s_2+8s_2^2)\sqrt{s_1(1-s_1)} 
+ (4s_2-2)(2s_1-1)\sqrt{s_2(1-s_2)} 
&= (4s-3)(1-4s)\sqrt{s(1-s)},
\label{conditionb} \\
s_1s_2 &= s^2.
\label{conditionc}
\end{align}
\end{subequations}}
\textcolor{black}{Substituting the 3rd condition in Eq.~(\ref{conditionc}) into Eqs. (\ref{conditiona}) and (\ref{conditionb}), reduces the problem into a quintic equation for $s_2$, \begin{equation}
    16s_2^5 + (16s-32)s_2^4+(16+40s-176s^2+128s^3)s_2^3+(-56s + 240s^2-448s^3+512s^4-256s^5)s_2^2 + (17s^2-48s^3+32s^4)s_2 - s^3 = 0. 
\end{equation} Solving the above equation numerically gives $s_2$ for a definite value of $s$, and one can obtain from Eq. (\ref{conditionc}), $s_1 = s^2/s_2$. Thus, the Parrondo sequence $A'A'B'B'B'B'..$ is periodic. Considering $s = 0.5, \gamma = 0, \delta= 0$, we get $s_1 = 0.8967902128$ and $s_2 = 0.278771998$, with phases $\gamma_1 = \gamma_2 = \delta_1 = \delta_2 = 0$. }
\textcolor{black}{For the sequence, $A'A'B'B'B'B'...$, the walk exhibits periodic dynamics with period 24. Fig.~\ref{f26} shows the probability of the walker at the initial site ($x = 0$) evolving under the Parrondo strategy, $A'A'A'A'A'A'...$, $B'B'B'B'B'B'...$, and $A'A'B'B'B'B'...$ on a 4-cycle graph with the former two exhibiting chaotic dynamics and the latter yielding an ordered quantum walk.}

\subsection{\textcolor{black}{Quantum circuit realization of the Parrondo-based cryptographic protocol using the sequence $A'A'B'B'B'B'...$: }}
\textcolor{black}{The circuit construction for this Parrondo sequence would be similar to the one described in Sec .~\ref {S32}. Here, we briefly outline the steps of the protocol with the Parrondo sequence, $A'A'B'B'B'B'...$}

    \textcolor{black}{\textbf{1. Generating a chaotic public key:} Alice (receiver) generates the public key using the chaotic operator, $B'$. The chaotic public key $\ket{\Phi_{PK}} = B'B'\ket{l}\ket{x}$, with $\ket{l}$ being the coin state and $\ket{x}$ being the position state of the walker.\\
    \textbf{2. Public key transfer:} Alice, after generating the public key, sends it to Bob. This is done by applying SWAP gates between Alice's and Bob's qubits.\\
    \textbf{3. Message Encryption: }To encode the message $k$, Bob applies the spatial translation $T_k$ operator (see, Eq. (\ref{e35})) to public key $\ket{\Phi_{PK}}$ with $T_k$ for different values of $k\in \{0,1,2,3\}$ in a $4$-cycle graph as given in Eq.~(\ref{e38}).\\
    \textbf{4. Message transfer: } After encoding the message, Bob transfers it to Alice, which is done again by the SWAP gates applied to Bob's and Alice's qubits.\\
    \textbf{5. Message Decryption.} Alice now decrypts the message by applying the Parrondo sequence $G = (A'A'B'B'B'B')^3AABB$ on $\ket{\Phi(k)}$ defined in Eq. (\ref{e34}) with $GB'B' = (A'A'B'B'B'B')^4) = I$.\\  We provide the corresponding Algorithm 6 (Fig.~\ref{a5}) for realizing the cryptographic protocol in a quantum circuit (see, Fig.~\ref{ABBAckt}).}

\begin{figure}[H]
\caption{\textbf{Algorithm 6}: Parrondo's Paradox cyclic QW based Cryptographic Protocol on a 4-cycle graph with Parrondo strategy $A'A'B'B'B'B'..$}
\label{a5}
\centering
\fbox{%
\begin{minipage}{0.95\columnwidth}
\small
\begin{algorithmic}[1]
\State \textbf{Input:} Pre-shared private key inputs: Coin operators $A'$, $B'$ generating chaotic DTQW dynamics, time steps $t$, initial coin $\ket{l}$ and position state $\ket{x}$ of walker
\State Design the circuit with 6 qubits: Alice 3 qubits $q_0q_1q_2$, Bob 3 qubits $q_3q_4q_5$; Position qubits: $q_0q_1$, $q_3q_4$, Coin qubits: $q_2$ and $q_5$ 
\Statex \textbf{Public Key Generation by Alice}
\State Initialize $q_0q_1q_2$
\State Apply QFT on position qubits of Alice $q_1q_0$ 
\State Apply SWAP on $q_1q_0$ to get the correct output 
\For{$i = 0$ to $2$}
    \State Perform the coin operation B' on $q_2$
    \State Perform the phase rotation $P(-\pi)$ on $q_0$  
    \State Perform the phase rotation $P(-\pi / 2)$ on $q_1$
    \If {$\ket{q_2}$ is in $\ket{1}$}
        \State Perform the phase rotation $P(\pi)$ on $q_1$
    \EndIf
\EndFor
\Statex\textbf{Public Key transfer from Alice to Bob}
\State Perform the SWAP gates to swap Alice's qubits with Bob's 
\Statex \textbf{ Message Encryption by Bob}
\State Encrypt the message via $T_k^D$ operator on Bob's position qubits $q_3q_4$
\Statex \textbf{Message transfer from Bob to Alice} 
\State Apply SWAP gates to swap Bob's qubits with Alice 
\Statex \textbf{Message decryption by Alice}
\For{$i = 0$ to $22$} 
    \If{$i \bmod 6 = 1$ or $i \bmod 6 = 2$} 
        \State Perform the coin operation $A'$ on $q_2$
    \Else
        \State Perform the coin operation $B'$ on $q_2$
    \EndIf
    \State Perform the phase rotation $P(-\pi)$ on $q_0$  
    \State Perform the phase rotation $P(-\pi / 2)$ on $q_1$
     \If {$\ket{q_2}$ is in $\ket{1}$}
        \State Perform the phase rotation $P(\pi)$ on $q_1$
    \EndIf
\EndFor
\State Apply SWAP on $q_1q_0$ to get the correct output 
\State Apply IQFT on position qubits $q_1q_0$
\State Alice measures her position qubits $q_1q_0$
    \end{algorithmic}
    \end{minipage}%
}
\end{figure}
\begin{figure}[H]
    \centering
    \includegraphics[width=1\linewidth]{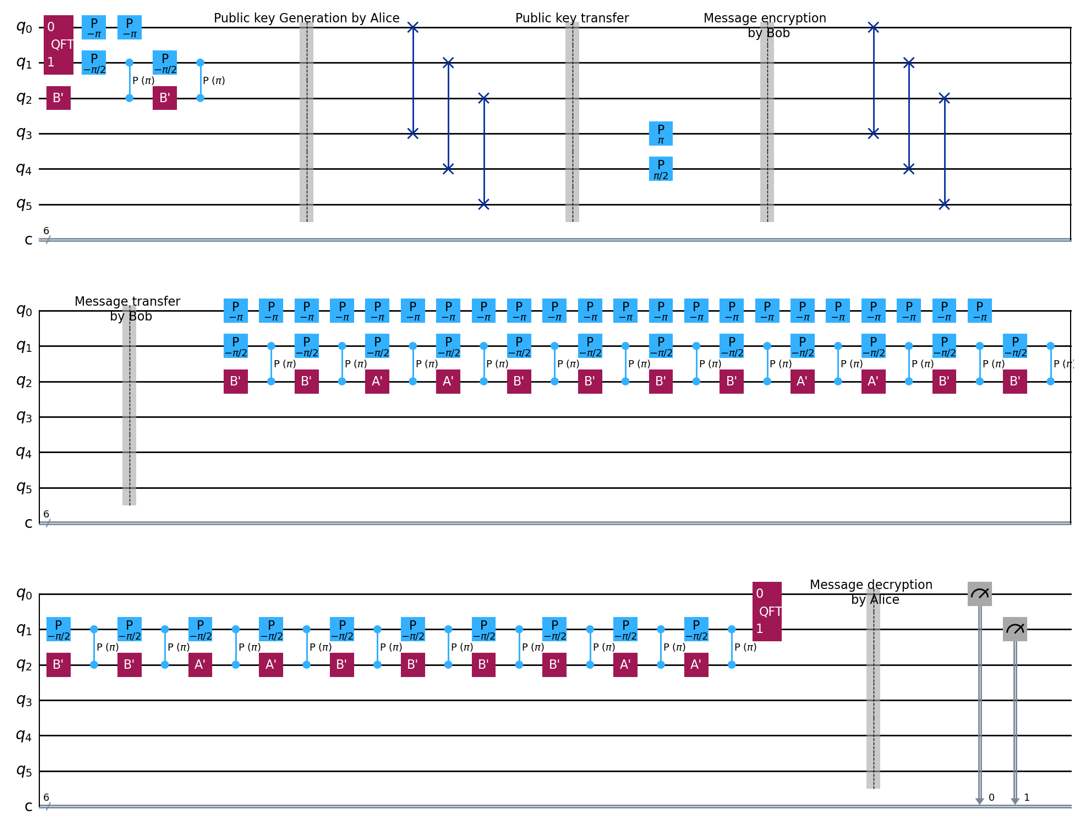}
    \caption{\textcolor{black}{Quantum circuit implementing the  Parrondo’s paradox–based cryptographic protocol with Parrondo sequence $A'A'B'B'B'B'...$. Qubits $q_0,q_1$ and $q_2$ represent Alice’s position and coin degrees of freedom, while qubits $q_3,q_4$ and $q_5$ correspond to Bob’s register. The circuit realizes public key generation by Alice with initial state $|000\rangle$, message $k = 1$ encoding by Bob, state transfer between the parties, and subsequent decryption performed by Alice.}}
    \label{ABBAckt}
\end{figure}
\begin{figure}[h!]
    \centering
    \includegraphics[width=0.8\linewidth]{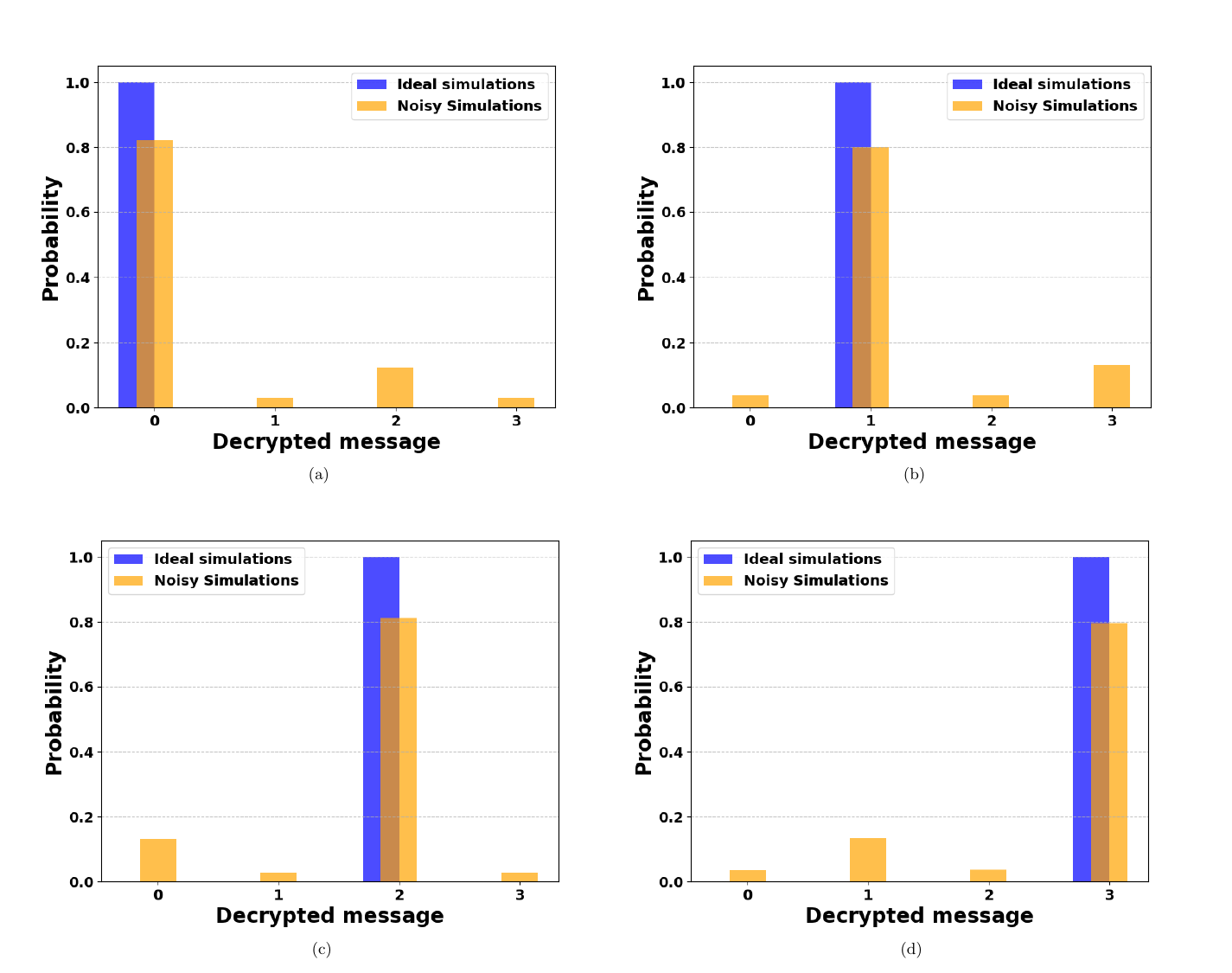}
    \caption{\textcolor{black}{Probability distribution for the Decrypted message, $k'$ for encoded messages (a) $k = 0 $, (b) $k = 1 $, (c) $k = 2$, and (d) $k = 3 $ with initial position $\ket{x} = |0\rangle$ such that $k'=k$ implemented in \texttt{qiskit\_aer} with depolarizing noise and without noise for $10^5$ shots using the $A'A'B'B'B'B'...$ Parrondo sequence on a 4-cycle graph. The peak at the correct message state in both ideal and noisy simulations demonstrates successful recovery of the encoded message, indicating that the periodic Parrondo dynamics required for decryption remain robust under realistic noise conditions.}}
    \label{f25}
\end{figure}
\begin{figure}[H]
    \centering
    \includegraphics[width=0.8\linewidth]{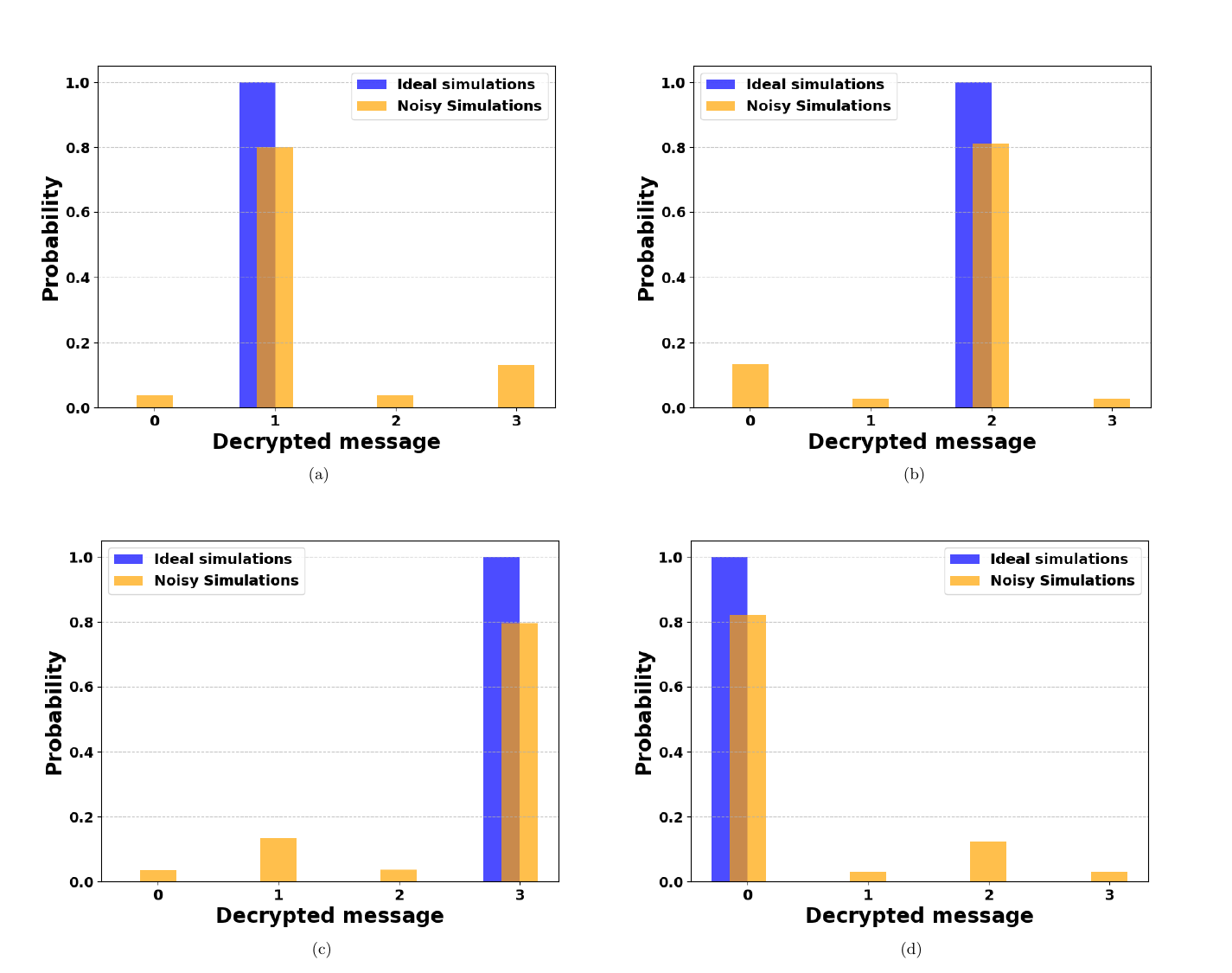}
    \caption{\textcolor{black}{Probability distribution for the Decrypted message, $k'$ for encoded message $k = 1$ with initial states (a) $\ket{x} = 0 $, (b) $\ket{x}  = 1 $, (c) $\ket{x}  = 2$, and (d) $\ket{x}  = 3 $  such that $k'=(k+x)\text{ mod } K$ implemented in \texttt{qiskit\_aer} with depolarizing noise and without noise for $10^5$ shots using the $A'A'B'B'B'B'...$ Parrondo sequence on a 4-cycle graph. The peak at the correct message state in both ideal and noisy simulations demonstrates successful recovery of the encoded message, indicating that the periodic Parrondo dynamics required for decryption remain robust under realistic noise conditions, irrespective of initial state.}}
    \label{f27}
\end{figure}
\begin{figure}[H]
    \centering
    \includegraphics[width=0.5\linewidth]{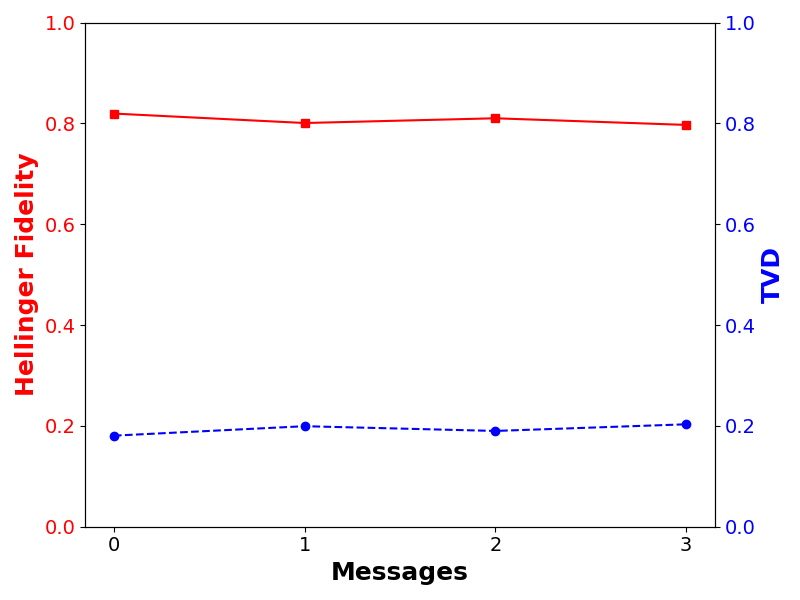}
    \caption{\textcolor{black}{Hellinger Fidelity (HF) and total variance distance (TVD) for different messages decrypted by Alice with initial position $\ket{x} = \ket{0}$ and coin $\ket{s} = \ket{0}$ implemented in \texttt{qiskit\_aer} with depolarizing noise and without noise for $10^5$ shots using the $A'A'B'B'B'B'...$ Parrondo sequence on a 4-cycle graph.}}
    \label{HF_new}
\end{figure}

\textcolor{black}{Fig.~\ref{f25} shows the probability distribution of the decrypted message obtained with initial state $\ket{l=0}\ket{x = 0}$. Numerical simulations of the circuit (see, Fig.~\ref{ABBAckt}) are carried out in \texttt{QISKIT} under both ideal and depolarizing noise conditions, showing behavior qualitatively consistent with that presented in Figs. \ref{f2}. The encoded message is recovered with high probability under ideal evolution. Fig. \ref{f27} shows that the messages encoded using different initial walker states can also be reliably decrypted, with higher fidelity. In the presence of noise, the distribution broadens due to accumulated errors, but the correct message remains identifiable with high fidelity ($\approx 80\%$, see Fig.~\ref{HF_new}). 
These results further support that the proposed protocol is not tied to a single specific sequence, but relies on the broader mechanism of periodicity emerging from deterministic combinations of coin operations.}


\twocolumngrid

\end{document}